\documentclass[12pt]{article}

\usepackage[top=1in, bottom=1.25in, left=1.20in, right=1.20in]{geometry}

\usepackage[american]{babel}
\usepackage[utf8x]{inputenc}
\usepackage{csquotes}
\usepackage{mathtools}
\usepackage{graphicx} 
\usepackage[natbibapa]{apacite}
\usepackage[colorlinks=true, urlcolor=blue, linkcolor=blue, citecolor=blue]{hyperref}
\usepackage{soul}
\usepackage{longtable}
\usepackage{threeparttablex} 
\usepackage{booktabs} 
\usepackage{pdflscape}
\usepackage{array}
\usepackage{multirow}
\usepackage[font=normalsize]{caption}
\usepackage{makecell} 
\usepackage[table]{xcolor} 
\usepackage{amsmath} 
\usepackage{bbold} 
\usepackage{xcolor} 

\newcommand\fnote[1]{\captionsetup{font=scriptsize}\caption*{#1}}
\newcommand\fsource[1]{\captionsetup{font=scriptsize, justification=justified, singlelinecheck=false, aboveskip = 2pt}\caption*{#1}}

\begin{document}

\begin{titlepage}
\title{Predicting Default Probabilities for Stress Tests: A Comparison of Models}
\author{Martin Guth\thanks{Oesterreichische Nationalbank (OeNB), Otto-Wagner-Platz 3, A-1090 Vienna, Austria. Email: martin.guth@oenb.at. The views expressed in this paper are those of the author and do not necessarily reflect those of the Eurosystem or the OeNB. The author would like to thank Florian Huber and Robert Ferstl for helpful comments that significantly improved the article.}}
\date{\today}
\maketitle
\begin{abstract}
\noindent Since the Great Financial Crisis (GFC), the use of stress tests as a tool for assessing the resilience of financial institutions to adverse financial and economic developments has increased significantly. One key part in such exercises is the translation of macroeconomic variables into default probabilities for credit risk by using macrofinancial linkage models. A key requirement for such models is that they should be able to properly detect signals from a wide array of macroeconomic variables in combination with a mostly short data sample. The aim of this paper is to compare a great number of different regression models to find the best performing credit risk model. We set up an estimation framework that allows us to systematically estimate and evaluate a large set of models within the same environment. Our results indicate that there are indeed better performing models than the current state-of-the-art model. Moreover, our comparison sheds light on other potential credit risk models, specifically highlighting the advantages of machine learning models and forecast combinations.\\
\vspace{0in}
\vspace{0in}

\bigskip
\end{abstract}
\setcounter{page}{0}
\thispagestyle{empty}
\end{titlepage}
\pagebreak \newpage

\section{Introduction}
\label{sec:intro}

Stress tests have a twenty-year history as tools for micro- and macroprudential supervision and are now used regularly by financial institutions and those who supervise them. The aim of such tests is to assess institutions' resilience to adverse financial and economic developments, as well as to contribute to the overall assessment of systemic risk in the financial system. In order to assess the behavior of financial institutions to stress, a multi-year macrofinancial scenario needs to be designed. It captures relevant systemic risks and the materialization thereof to generate stress in the system.

The present study focuses on the use of macrofinancial linkage models to translate the country-level scenario into bank-level risk parameters, which are an essential input to every stress test. These so-called satellite models thus provide scenario-conditional forecasts for the probabilities of default (PD). What is key in these models is the proper detection of signals from an array of \textit{global} variables which are not directly linked to the financial health of companies. 

Outside the context of stress testing, such credit risk models have a long-standing history  \citep{keeton:1987,wilson:1998} and still represent a very active research area in which various models in different setups have been proposed. Specifically, the literature covers linear models (see, e.g., \citealp{aver:2008}; \citealp{bofondi:2011}), vector autoregressive (VAR) models (see, e.g., \citealp{gambera:2000}; \citealp{pesaran:2006}; \citealp{castren2010}), panel models (see, e.g., \citealp{pesola:2001}; \citealp{castro:2013}), latent factor models (see, e.g., \citealp{koopman:2005}; \citealp{kerbl:2011}), quantile regressions \citep{schechtman:2012} and machine learning models \citep{jacobs:2018}. Interestingly, besides \citet{jacobs:2018}, in the machine learning literature, there seems to be little to no coverage of credit risk models that estimate or predict PDs. Even in very recent literature surveys that specifically cover specifically machine learning in banking risk management \citep{leo:2019} and machine learning in credit risk \citep{breeden:2020}, there seems to be just a few papers in the approximate vicinity of this topic.

The current state-of-the art satellite model for PD translation is Bayesian model averaging (BMA) \citep{raftery:1995}. It has a long track record as being a reliable tool for generating scenario-conditional projections for credit risk and is being adopted by more and more central banks and institutions. The inherent advantage of BMA is the explicit tackling of model uncertainty by operating on a large pool of competing models which are weighted by their predictive performance and combined to one final model. However, with easier access to sophisticated regression approaches provided by open source programming languages such as R \citep{R:2020}, Julia \citep{bezanson:2017} or Python \citep{vanrossum:2009} and the advent of new predictive models in the field of machine learning, the question arises if there are other models that could deliver better results. 


The aim of this paper is to conduct a systematic forecast comparison with a large number of different regression models to find the best performing credit risk satellite model. The winning model is evaluated for the ability to precisely forecast default probabilities conditional on a standardized set of macroeconomic variables as provided to financial institutions by the \citet{esrb:2020} for the EU-wide banking sector stress test. We implemented a total of 43 models, which can be assigned to 9 categories, ranging from conventional statistical models to more recent machine learning methods. We tried to encompass as many models as possible with a proven track record in forecasting linear and non-linear relationships and which were readily available within an R library \citep{R:2020}. Additionally, we also combine the models with different forecast combination approaches to further gauge their potential accuracy. For the purpose of this paper, we implement a framework that allows us to conduct this comparison with a standardized data set for all models, to tune the respective hyperparameters for each model and to cross-validate the results based on recursive pseudo out-of-sample forecasts.

There are only two papers in the literature that are related to our work. \citet{papadopoulos:2016} created a composite satellite model for stress testing by weighting candidate models from the full space of all possible variable combinations. \citet{grundke:2019} used a combination of BMA to select relevant variables and OLS to regress the selected variables onto a credit default index. They further analyzed different modifications to various steps in their estimation framework and assessed the different outcomes in terms of out-of-sample default rate forecasts. Both papers focus on the proper identification of the model given one estimation technique but lack comparisons across different procedures. 


Our paper contributes to the literature in the following ways: To the best of our knowledge, it represents the first systematic model comparison in the field of credit risk and stress testing. First and foremost, we deliver insights on a large number of potential credit risk satellite models across a wide range of modelling techniques. Since it is not evident which modelling technique will achieve the best results, we take a naive approach by testing many different models without prejudice on how well they will fare. Our results indicate that there are better performing models than the current state-of-the art model (i.e. BMA), which have not been mentioned in the literature until now. Second, due to the large variation in models, we are able to shed light on the potential positive effects of machine learning models and thus extend the scarce literature in this area. Moreover, the use of different forecast combination techniques across different sets of models shows that not only can these techniques help to hedge against model uncertainty, but they can also enhance the predictive accuracy.



The remainder of the paper is structured as follows. In Section~\ref{sec:setup} our estimation and evaluation framework are explained in detail, including the used data, models, hyperparameter tuning and performance measures. Thereafter, the results are shown in Section~\ref{sec:results}, which also presents a deep dive into the winning model. Section~\ref{sec:conclusio} concludes the paper.

\section{Design of the Forecasting Exercise}
\label{sec:setup}

This section outlines the setup of our forecasting exercise. First, we present the underlying data set and the applied transformations. Second, we give a short summary of the 43 models within the 9 overarching categories. Third, as many models need a prior setting of parameters, we discuss the process of tuning the respective hyperparameters. Fourth, we briefly discuss the measure used to evaluate the forecasting performance.

\subsection{Data}
\label{sec:data}

The data involved in the modelling exercise refer to Austria and include as dependent variable a measure for the probability of default for the non-financial corporate sector and macroeconomic and financial data as independent variables. The deployed default probabilities are based on Moody’s KMV Expected Default Frequency (EDF) measure, available at quarterly frequency from 2002Q2 to 2019Q2 ($T = 1,\dots,69$). The EDFs have been successfully deployed in multiple credit risk models (see, e.g. \citealp{alves:2005}; \citealp{castren2010}; \citealp{gross:2019}). In adherence to the IMF's approach of credit risk modelling in their Financial Sector Assessment Programs (FSAP), we also take the provided mean PD measures and not the median (see \citealp{imf:2020}). Hence, the measure relates to the average default probability across non-financial corporations. 

The independent covariates are based on the variables within the macroeconomic scenario as designed for the EU-wide stress tests by the European Systemic Risk Board \citep{esrb:2020}. These variables are real GDP growth (\textit{GDP}), the unemployment rate (\textit{UNE}), the inflation rate (\textit{INF}), real estate price growth (\textit{RRE}), stock price growth (\textit{EQP}), exchange rates (\textit{EXR}) and short-term (\textit{STR}) and long-term interest rates (\textit{LTR}). However, the actual scenario figures are not needed for the estimation as we conduct pseudo out-of-sample forecasts for which the actual default probability time series is needed. Hence, the scenario serves as guidance for the choice of covariates, but the models are evaluated on the basis of historical figures. 



Although the range of the time series is limited by the availability of the default probability, it still includes non-linear events such as the financial crisis of 2008 and the European sovereign debt crisis of 2011, which is an important feature for credit risk satellite models, as they must be able to estimate and predict such structural breaks. We get data on real GDP, Harmonized Index of Consumer Prices (HICP) and the unemployment rate from Eurostat. The GDP figure is seasonally and calendar adjusted and transformed to year-on-year (YoY) growth rates to fit the figures used by the ESRB. Similarly, the HICP is also transformed to YoY growth rates to match the ECB definition of inflation \citep{ecb:HICP}. The real estate prices and EUR/USD exchange rate are sourced from the ECB's Statistical Data Warehouse (SDW). The house prices cover all new and existing residential properties across all dwelling types and are also transformed into YoY growth rates. Finally, we take 10-year government bond yields as long-term interest rates, 3-month Euribor as short-term interest rates and the equity price index for Austria, the ATX, from the OECD database. The equity price index is also transformed into YoY growth rates.

\begin{figure}[ht!]
    \centering
    \includegraphics[width=1\textwidth]{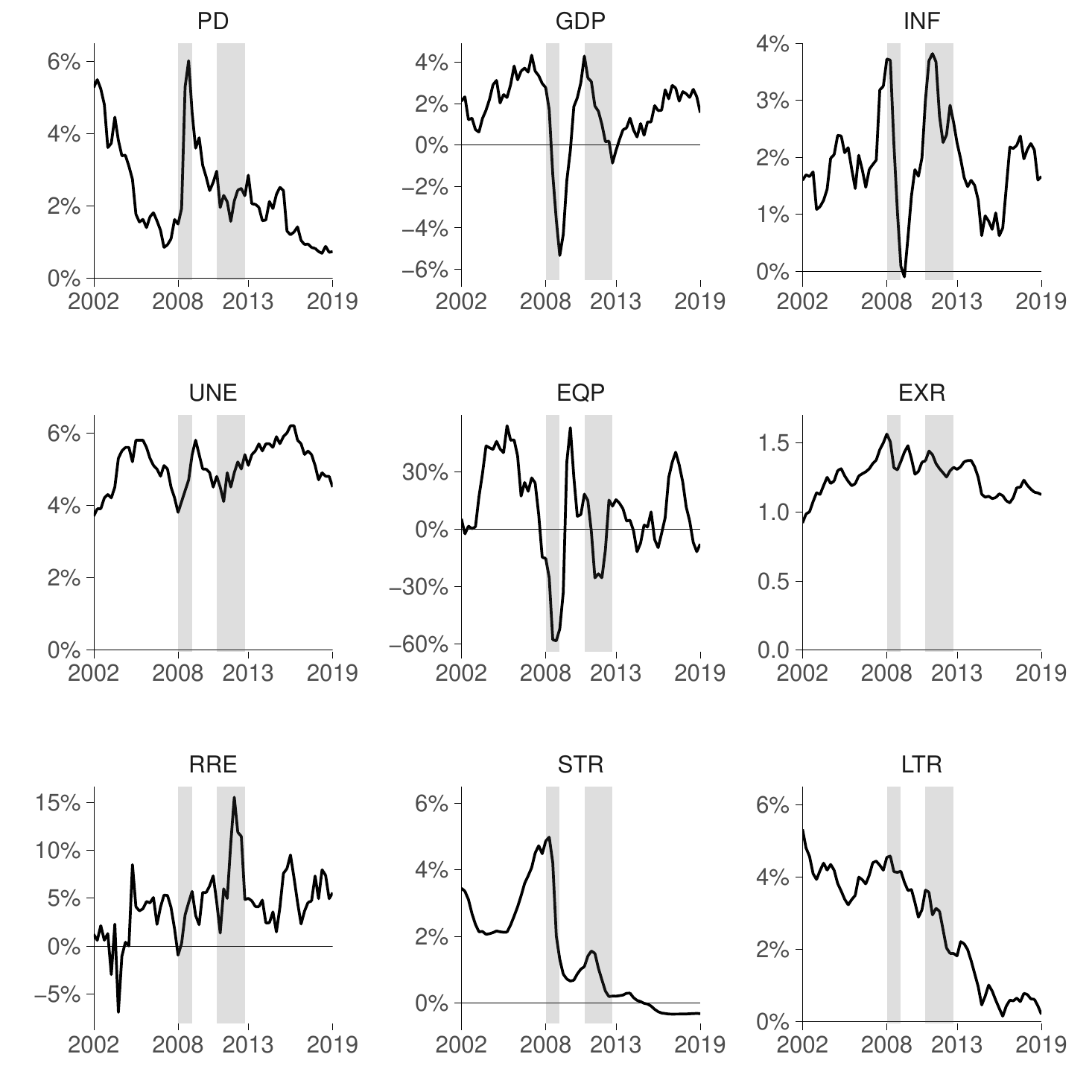}
    \caption{Overview variables and recession indicators}
    \fnote{Note: the grey shaded areas mark Austria's economic crisis periods based on the recession indicator of the Federal Reserve Bank of St. Louis. These two areas are the Great Financial Crisis (2008 Q2 - 2009 Q2) and the European sovereign debt crisis (2011 Q1 - 2013 Q1).}
    \fsource{Source: Moody’s KMV EDF, Eurostat, ECB SDW, OECD, Federal Reserve Economic Data (FRED).}
    \label{fig:data}
\end{figure}

A first descriptive analysis of the variables and potential correlations among them can be drawn from Figure~\ref{fig:data}. The chart depicts the variables without further transformations and the grey shaded areas mark the two previous crisis periods -- the GFC (2008 Q2 - 2009 Q2) and the European sovereign debt crisis (2011 Q1 - 2013 Q1). The sample starts exactly at the peak of the crisis that unfolded in the aftermath of the burst of the dotcom bubble in 2000, the uncertainty triggered by the 9/11 attacks and the very volatile years after the introduction of the euro as a new currency from 2000 - 2001. The economic shockwave led to a significant increase in expected and actual corporate defaults around the globe \citep{altman:2003}. 

In the case of Austria, the default probability starts with a value of 5.3\% and decmidrules until the start of the financial crisis, when it reaches the sample maximum of 6\%. The pattern during the European sovereign debt crisis is not as clear, even though there is a small increase in the EDFs towards the end of 2012. The movements of the macro and financial variables behave as expected during the downturns. In both periods, we see significant drops in GDP growth and inflation, an uptick in the unemployment rate, large negative distortions on the stock market, devaluations of the euro vis-a-vis the US dollar and an increase in real estate prices reflecting a flight to safe investments, and we also see the ECB reacting to these events by pushing the short-term interest rate and indirectly the long-term interest rates towards zero and beyond. Even though the co-movements of the variables within the structural breaks seem to be going in sensible directions (e.g. GDP down, PD up), the credit risk satellite models in our setup will need to be able to pick up correct signals in calmer periods. In the majority of European countries, the same co-movement patterns have been observed in the last two decades. Therefore, our conclusions based on Austria can be generalized to other regions. 


As a last transformative step, we need to make sure that the variables have optimal properties and behave well in predictions. First, to ensure that the point forecasts of the default probability is bounded between a 0\% and 100\% interval, we apply the following logit transformation for the regression and calculation of the performance measures,

\begin{equation}
y = \mathrm{logit}(PD) = \mathrm{log} \left( \frac{PD}{100 - PD} \right)
\end{equation}

\noindent Second, we analyze the trend and cyclical components of the variables and perform a series of unit root tests. The seasonal decomposition by loess \citep{robert:1990} shows that all variables, except GDP (since it has already been adjusted), exhibit a form of cyclicality, which we remove in due course. In order to get a clear picture of the stationarity of the variables we deploy the augmented Dickey-Fuller (ADF) test \citep{dickey:1979}, the Elliott, Rothenberg \& Stock (ERS) test \citep{elliott:1996}, the Phillips-Perron (PP) test \citep{phillips:1988} and the Kwiatkowski-Phillips-Schmidt-Shin (KPSS) test \citep{kwiatkowski:1992}. Without further transformations, all variables would suffer from unit roots. Hence, by taking the first difference for each variable, the test statistics in each unit root test indicate robust level and trend stationarity.

\subsection{Econometric Models}
\label{sec:models}

In this section we give a short overview of, and introduction to, the adopted credit risk satellite models. In order not to artificially expand the paper, the formal definitions of the models are not mentioned below. The interested reader is referred to the publications cited in each paragraph. The selection of the models was based, on the one hand, on the desire to cover as many models as possible with a wide array of different features. In doing so, we want to extend the existing model space in the literature from mainly linear models to non-linear, data-driven models with a special focus on regularization. Especially the latter point will turn out to be very important when analyzing the results. On the other hand, as the implementation of so many models is time-intensive, we focus our attention on proven models that are readily available, using open source computing environments, such as R \citep{R:2020}.\footnote{For a compact overview of the models and R libraries see Table~\ref{table:models} in the annex.}

In total, we have implemented 43 models which are placed in 9 overarching groups to give the reader a better overview of the models at hand. All satellites use the same basic model structure in which the dependent variable $\mathbf{y} = (y_1,\dots,y_T)'$ is described as a function of contemporaneous and lagged predictors $\mathbf{X} = (x_1,\dots,x_T)'$, such that

\begin{equation}
\label{eq:model}
    \mathbf{y} = f(\mathbf{X}_t, \mathbf{X}_{t-1},\dots,\mathbf{X}_{t-p}) + \varepsilon 
\end{equation}


\noindent The actual approximation of $f()$ is dependent on the respective model in our comparison. Notice that the equation is ordered by the lags, i.e. $p = 0, 1,\dots P$. This fact is important for certain stepwise regressions and it achieves overall better prediction results than with the equation being sorted by the variables. We set $P=4$ as a standard choice for quarterly data. We follow \citet{gross:2019} by forcing a "closed" lag structure without gaps between the first and fourth lag for the initial equation. \citet{gross:2019} found in their analysis that this type of structure appeared to be more meaningful and robust. However, models with regularization or variable selection are not constrained in their choice of the proper equation. 

We do deviate from some papers in the literature on credit risk satellite models in that we do not include an autoregressive lag of the dependent variable. When we tested both options, nearly all models showed a worse performance with the autoregressive lag than without, while the overall ranking across the models remained stable. Moreover, we did not want to increase the already large set of contemporaneous and lagged regressors ($n = 40$) in combination with the limited sample length ($T = 64$).

\vspace{5mm}
\noindent \textbf{(Generalized) Linear Models}
\vspace{5mm}

\noindent The first class of competitors are (generalized) linear models with a proven track record in a number of disciplines. We start with naive benchmarks based on a standard ordinary least squares (OLS) regression (short name: \textbf{lm}) and its robust alternative based on the M-estimator (\textbf{rlm}) as defined in \citet{huber:1992}.


Nevertheless, due to the high number of variables relative to the sample, overfitting in the context of predictions is problematic. Thus, the next group of models uses regularization techniques to reduce the number of covariates. More precisely, we implement a forward selection algorithm (\textbf{lmfs}) \citep{hocking:1976} and a least-angle regression (\textbf{lars}) by \citet{efron:2004}.\footnote{The corresponding backward selection procedure yields exactly the same best model subset. Using the Akaike Information Criterion (AIC) instead of the BIC, diminishes the out-sample performance as too many predictors are chosen as being relevant.} 


Shrinkage estimators deviate from the stepwise approach as they penalize the coefficients to reduce multicollinearity in the equation. Here we chose the ridge estimator (\textbf{ridge}) (\citealp{tikhonov:1943}; \citealp{phillips:1962}) and, as the ridge cannot yield sparse equations because no covariates are dropped due to the regularization procedure, the least absolute shrinkage and selection operator (\textbf{lasso}) by \citet{tibshirani:1996}. 


We further implement three refinements to the lasso estimator that try to deal with certain shortcomings of the original estimator. First, if one takes the temporal structure of the data into account, we get the fused lasso (\textbf{flasso}) by \citet{tibshirani:2005}, which adds a second penalty to the differences of the coefficients. Second, \citet{meinshausen:2006} showed that using only one penalty factor implies an inherent conflict between model selection and shrinkage estimation, leading to many noise variables in the final variable set. Thus, \citet{meinshausen:2007} introduced the relaxed lasso (\textbf{relaxo}) with a new parameter controlling the applied shrinkage. Third, lasso introduces a potential bias for large coefficients \citep{fan:2001}, which can be tackled by adding weights to the regularization term -- i.e. the adaptive lasso (\textbf{adalasso}) by \citet{zou:2006}.

Building on the strengths and weaknesses of the ridge and lasso estimator and combining both penalties, \citet{zou:2005} introduced the elastic net (\textbf{glmnet}). This algorithm includes a complexity parameter controlling the strength of the regularization and a mixing parameter between ridge and lasso regression. 


The next three models also belong in the class of shrinkage estimators, yet they do not alter the coefficients but the covariates themselves. Specifically, they assume that the variables can be described as a linear combination of a reduced set of factors and loadings. These are principal component regressions (\textbf{pcr}), independent component regression (\textbf{icr}) by \citet{comon:1994} and partial least squares (\textbf{pls}) by \citet{wold:2001}.


Until now the models assumed that the estimated parameters are fixed and driven by an unknown underlying data-generating process. With the upcoming models we want to venture into Bayesian territory and thereby assume that the parameters are random variables following certain distributions for which we can apply prior knowledge. As before, starting with a simple alteration of the linear model in which the coefficients follow a Student-t distribution (\textbf{bayesglm}) \citep{gelman:2008}. Additionally, we introduce shrinkage via the Bayesian ridge regression (\textbf{bridge}) and the spike-and-slab (\textbf{spikeslab}) prior \citep{ishwaran:2005}.

Unsurprisingly, there is also a Bayesian alternative for the lasso estimator (\textbf{blasso}) introduced by \citet{park:2008}. The results are very similar to the original lasso algorithm, but the Bayesian treatment has the advantage that the penalty factor has not to be determined via cross-validation but can be implicitly derived in a fully Bayesian fashion. To further reduce the time-intensive and computationally demanding calculations, \citet{cai:2011} combined an empirical Bayes (EB) method with lasso to create the empirical Bayesian lasso (\textbf{eblasso}).

As a last subgroup of the Bayesian models, we introduce "global-local" shrinkage estimators that, as the name suggests, introduce a global shrinkage parameter pushing the coefficients uniformly towards the origin, while the local parameter allows for coefficient-specific deviations. Although there are many different priors to choose from, we settled for the Dirichlet–Laplace prior (\textbf{dlbayes}) by \citet{bhattacharya:2015}, the horseshoe prior (\textbf{horseshoe}) by \citet{carvalho:2010} and the extended horseshoe prior by \citet{bhadra:2017} named horseshoe+ (\textbf{horseshoePlus}), which exhibits significant improvements in the case of "ultra-sparse" signals (i.e. nearly all coefficients are zero).


Now coming to the last subgroup of the (generalized) linear models: ensembles of linear models. We deploy a straightforward gradient boosting algorithm (\textbf{bstlm}), which is based on linear models as weak learners.


\vspace{5mm}
\noindent \textbf{Model Averaging}
\vspace{5mm}

\noindent Variable selection, either in the way of regularization or shrinkage, can lead to an over-correction due to too many variables being penalized and thus biased estimates of the remaining covariates and too narrow confidence intervals as the inherent model uncertainty is not taken into account by focusing on only one equation \citep{lukacs:2010}. Therefore, we introduce three model averaging models that can overcome these issues by combining multiple equations of the same base model. First, Mallow’s model averaging (\textbf{mma}) by \citet{hansen:2007}, which forms the final model by weighting multiple nested models based on minimized mean squared forecasting errors. Second, jackknife model averaging (\textbf{jma}), introduced by \citet{hansen:2012}, calculates the weights by minimizing the leave-one-out (LOO) cross-validated residuals. Third, we introduce the current state-of-the-art credit risk satellite model: Bayesian model averaging (\textbf{bma}) \citep{raftery:1995}. With this technique, the weights are based on Bayes’ theorem and the posterior model probabilities thereof. 


\vspace{5mm}
\noindent \textbf{Exponential Smoothing}
\vspace{5mm}

\noindent Forecasts based on exponential smoothing have a long and successful track record going back to the late 1950s \citep{brown:1957}. We implement a simple exponential smoothing (\textbf{es}) algorithm with additive errors and no trend or seasonal component based on \citet{hyndman:2008}.\footnote{The model type for error, trend and seasonality -- in our case ANN -- is implicitly chosen by the sample-size-corrected AIC. This is inline with our data transformations, which include de-seasonalizing and differencing. The exogenous variables are also chosen based on this criterion.} As the choice of model types is crucial for the performance of the model, \citet{svetunkov:2016} introduced complex exponential smoothing (\textbf{ces}), which avoids the artificial distinction of a time series in level, trend, and seasonality.

\vspace{5mm}
\noindent \textbf{(Generalized) Additive Models}
\vspace{5mm}

\noindent The class of generalized additive models (GAMs) marks the point from which the presented models become more non-parametric and data-driven in their estimation routines. A feature which is often attributed to machine learning models.

GAMs allow more flexibility compared to a standard linear model due to the built-in smoothing function. Additive models were originally proposed by \citet{friedman:1981}, and the first model in this class will be the one which they proposed in their seminal paper, namely Projection Pursuit Regression (\textbf{ppr}). Furthermore, we implement boosted smoothing spline (\textbf{bstpline}), which utilizes the same gradient boosting algorithm as the linear version (i.e. bstlm). Lastly, a boosted generalized additive model (\textbf{boostgam}) as outlined by \citet{schmid:2008} is implemented, combining the features of the first and second models.


\vspace{5mm}
\noindent \textbf{Multivariate Adaptive Regression Splines}
\vspace{5mm}

\noindent Another regression technique, which was proposed by Jerome Friedman, is the multivariate adaptive regression spline (\textbf{mars}) \citep{friedman:1991}. Somewhat similarly to (generalized) additive models, the setup is now fully non-parametric, consisting of linear combinations of hinge functions. This extension of linear models allows the implicit modelling of non-linearities and interactions between variables.

\vspace{5mm}
\noindent \textbf{Support Vector Machines}
\vspace{5mm}

\noindent Support vector machines (SVM) have a long-standing history in providing robust classifications for linear and non-linear problems. The work of \citet{drucker:1997} introduced the concept of support vectors to regression problems. For our model comparison, we utilize a specific SVM model called a L2-regularized L1-loss support vector regression (\textbf{svr}), which uses a linear kernel and, as the name states, performs ridge regularization on the covariates \citep{hsieh:2008}. 

The second model in this class is a relevance vector machine (RVM) which has the same functional form as SVMs but uses Bayesian inference to estimate the equation \citep{tipping:2001}. The advantages are that, compared to SVMs, RVMs are highly sparse and no prior cross-validation is needed to tune the cost function. We deploy the model with the standard Gaussian radial basis kernel (\textbf{rvm}), which delivered the best predictive performance.

\vspace{5mm}
\noindent \textbf{Gaussian Process}
\vspace{5mm}

\noindent A Gaussian process (GP) regression \citep{williams:1996} is a Bayesian kernel machine which bridges the gap between Bayesian linear models or spline models and SVMs. A GP builds on a covariance matrix which is used to assess how much information contiguous observations convey about each other. By applying a kernel matrix, in our case a polynomial kernel (\textbf{gp}), the estimation is extended into a non-linear realm \citep{karatzoglou:2006}. 

\vspace{5mm}
\noindent \textbf{Tree-Based Models}
\vspace{5mm}

\noindent Similar to the class of (generalized) linear models, tree-based models also contain a multitude of different approaches, ranging from simple to more complex. The first subclass are rule-based tree structures such as classification and regression trees (\textbf{cart}) \citep{breiman:1984} that recursively split the data set to make predictions about the outcome variable. As a further advancement in this field, \citet{hothorn:2006} introduced conditional inference trees (\textbf{ctree}) which tackle the issues of overfitting and selection bias prevalent in CART by introducing permutation tests during the partitioning.

In the realm of tree-based models, ensemble learning methods are more common than in the case of linear models. One of the most well-known ensemble models is random forests (\textbf{rf}) by \citet{breiman:2001}. The method combines multiple decision trees which are trained on different subsampled parts of the data and with random subsets of variables. The results of all trees are averaged across to form the final prediction (i.e. bootstrap aggregation or bagging).\footnote{We want to note that the standard implementation of bagging in the used R library \texttt{ranger} is not sensitive to time series data as it assumes iid data. Currently, there exists no available library that includes bootstrap methods for dependent data.} The low bias in the results comes at the price of large variance. In order to tackle these issues, \citet{geurts:2006} suggested extremely randomized trees (\textbf{ert}) which use the whole data set for each tree instead of subsampling and randomize the splitting rule for each node.

Another popular modelling technique is gradient boosting \citep{friedman:2001}, which is again an ensemble method, but in contrast to random forests, gradient boosted trees (or gradient boosting machines, GBM) are built sequentially. Each new tree improves the shortcomings of former trees, combining the results along the way. We opt for a standard Gaussian loss function (\textbf{bsttree}). It may come as no surprise that there also exists a Bayesian version of GBM called Bayesian additive regression trees (\textbf{bart}) established by \citet{chipman:2010}. Instead of combing the trees via a learning rate, an iterative backfitting Markov chain Monte Carlo (MCMC) algorithm is used. As we will see in Section~\ref{sec:results}, this flexible setup turns out to be the winning model.

The last competitor in this class of models is called node harvest \citep{meinshausen:2010} and settles itself between easy-to-understand regression trees and more accurate ensembles like random forests (\textbf{enstree}). The model delivers sparse results by initially creating random nodes and then finding suitable weights for each node based on an empirical loss function.

\vspace{5mm}
\noindent \textbf{Neural Networks}
\vspace{5mm}

\noindent The last class of competitor models is one of the earliest and most commonly used techniques in machine learning: neural networks \citep{mcculloch:1943}. As a first model, we deploy a deep neural network (\textbf{nn}) with three hidden layers using resilient backpropagation with weight backtracking \citep{riedmiller:1994}. The last model will be again Bayesian, namely a Bayesian regularized neural network (\textbf{brnn}). This model fits a two-layer neural network as described in \citet{foresee:1997}. In contrast to classical neural networks, Bayesian networks are graphical models in which each node represents a variable with probabilistic relationships among them.

\subsection{Hyperparameter Tuning}
\label{sec:tuning}

The increasing complexity of the models outlined above is also reflected in the number of parameters that need to be set before a model can be estimated. These hyperparameters control complexity and are thus crucial ingredients to the overall outcome of the model. Although many machine learning libraries provide default values for most parameters, \citet{olson:2018} showed that tuned hyperparameters can significantly reduce the variance compared to the out-of-the-box values. 

There are different methods to find the most suitable set of hyperparameters (see \citealp{feurer:2019} for an overview), again with different layers of complexity. We choose a model-free, non-black-box method for this task: grid search. After specifying a set of values for each parameter, grid search evaluates each set combination. The drawback of grid search is the possible large number of combinations that must be evaluated. However, to tackle this problem, we combine grid search with expert judgment. Specifically, on the one hand, we conduct research on the proper parameter space for each model and, on the other hand, we manually fine tune the grid to reduce the computational burden. In combination with a highly efficient implementation by the \texttt{caret} library \citep{kuhn:2020}, the whole process stays feasible and very transparent.\footnote{We implement our own grid search for models which are not implemented in \texttt{caret}.}

For the performance measure we follow \citet{hyndman:2006} and stay within the realm of the well-known scale-dependent measures. The used data set stays the same across the models and we thus do not need to take the possible different data characteristics in consideration. Therefore, we choose the mean absolute error (MAE) as our indicator on how well a model predicts the default probability. Since some models exhibit an unstable forecasting behavior, we disregarded the more widely used mean square error (MSE) or root mean square error (RMSE) as they are more sensitive to outliers \citep{armstrong:2001}.\footnote{For completeness' sake, we also implemented the RMSE, mean absolute percentage error (MAPE) and mean absolute scaled error (MASE). However, the tuning parameters and model rankings are nearly identical.} 

Finally, in order to generalize the parameters for different sample lengths and time periods, a cross-validation strategy is introduced. More precisely, we apply a rolling-origin evaluation, starting with an initial training set of $T_{t,1} = 1,\dots,41$ and a fixed test (or holdout set) of $T_h = 12$, representing the 12 quarters we want to forecast and base our model comparison upon. In each iteration of the cross-validation, the training set is extended by one observation, leading to a total of 12 estimation rounds and a final training set of $T_{t,12} = 1,\dots,52$.\footnote{Due to the limited data availability and the large set of predictors, we do not deploy a fixed, moving window.} The best performing sets of hyperparameters are saved and used in the following estimations. The fairly large initial training set is justified by our sizeable set of $n = 40$ predictors, which allow us to estimate each model without a constraint on their ability to restrict the set of predictors with regularization. As a result, to obtain the hyperparameters for 37 models, around 15,000 estimations have to be carried out.\footnote{An overview of the tuned hyperparameters per model can be found in Table~\ref{table:models}.}

\section{A Comparison of Forecasts}
\label{sec:results}

For the same reasons as outlined above, the estimation of the 43 models is carried out by deploying the same cross-validation strategy. The only difference is that the initial training set is $T_{t,1} = 1,\dots,4$. Hence, we provide for the initial estimation only one full year of data to the models. In combination with the large set of variables, this enables us to gauge the performance of the models under such extreme overfitting conditions. Using such as small data set also reflects reality, in which data availability is often limited to incomplete data or only a few recent observations. Given these circumstances, we would assume that especially regularized models and machine learning models, which are often advertised for their ability to handle such cases well \citep{bzdok:2018}, will fare well in the comparison.

As with the hyperparameter tuning, the MAE is again used as our main evaluation factor. After each cross-validation window, the pseudo out-of-sample forecast is computed and compared to the actual PD time series. We thereby follow the idea outlined by \citet{hastie:2017} that for the purpose of model comparisons based on conditional forecasts, we do not need to focus on the causal relationship between the variables but identify statistical dependencies that are stable over time. Thus, this section will not present any results on the estimated coefficients, but only focus on the forecasting performance. 

Before we come to the evaluation of the models in terms of their predictive accuracy, we need to make sure the comparison is as fair as possible. Specifically, if models exhibit certain instabilities during the estimation or deliver no proper output at all, combining these predictions in one final score can lead to distorting effects on the overall ranking of the models.\footnote{In the majority of cases the instabilities occurred due to over-regularization effects, which led to flat forecasts or algorithms not being estimable in the case of $T \ll n$.} Thus, if 25\% of the predictions of a model fall in such a category, the whole model is dropped for further processing. As a consequence, 21 models have been dropped in due process -- 16 of which due to estimation instabilities and 5 could not estimate Eq.~\eqref{eq:model} in the case of $T \ll n$. Due to this process, the multivariate adaptive regression splines model has been dropped and thus the whole category, leaving 22 models in eight categories.

\begin{figure}[t]
    \centering
    \includegraphics[width=1\textwidth]{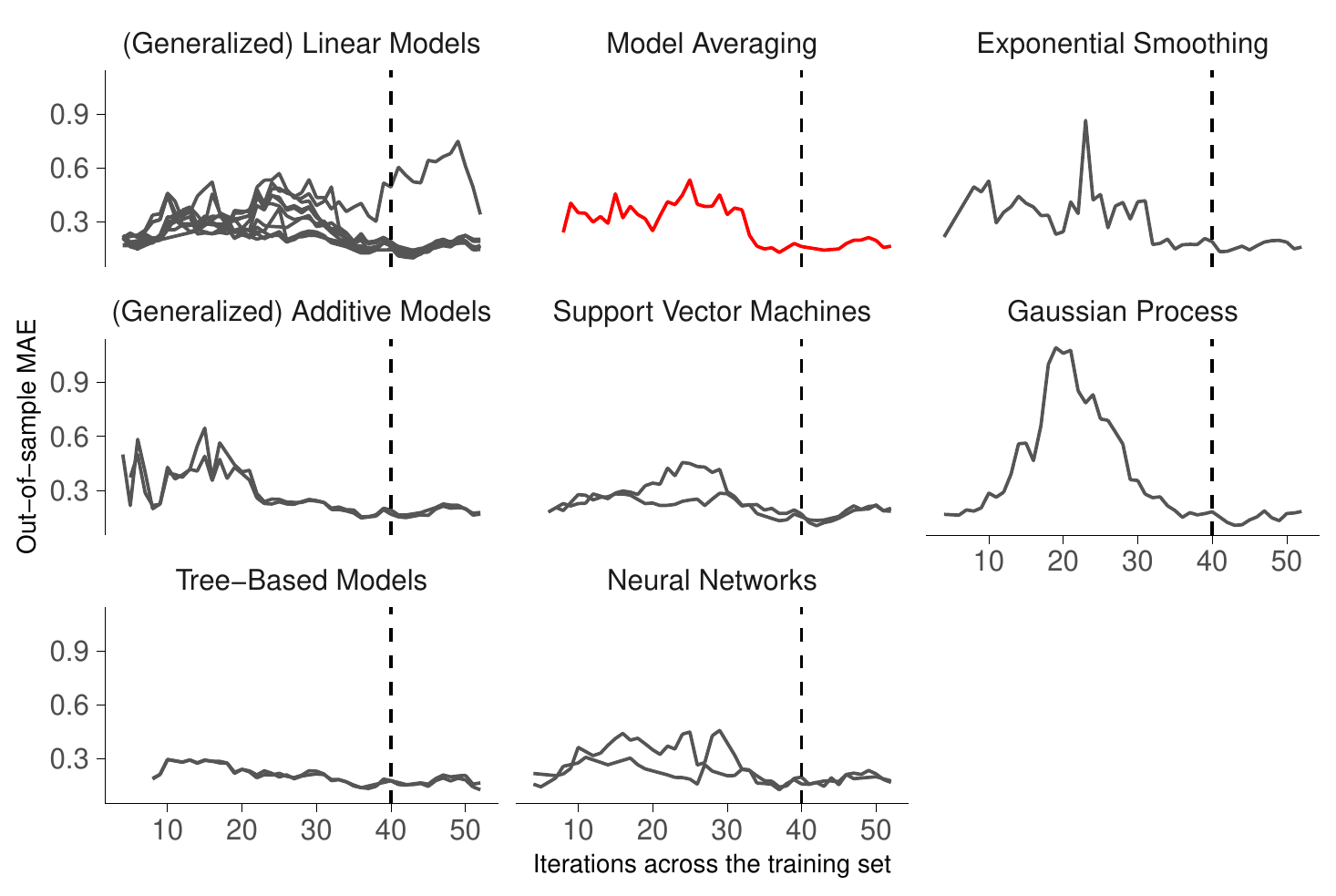}
    \caption{Evolution of out-of-sample MAE}
    \fnote{Note: The chart depicts the Mean Absolute Error (MAE) for the remaining 28 models across the 9 categories. The performance criterion is calculated for each cross-validation window. The dashed vertical lines indicate the overfitting threshold after which the sample $T$ is larger than the number of covariates $n$. The red line depicts the current state-of-the-art model Bayesian Model Averaging (\textbf{bma}).}
    \label{fig:linechart}
\end{figure}

Figure~\ref{fig:linechart} plots the evolution of the out-of-sample performance criterion across all cross-validation windows. The dashed vertical lines indicate the overfitting threshold after which sample $T$ is larger than the number of covariates $n$. At first glance, the figure indicates a significant variation of prediction accuracy across model categories, but also within the groups. Another key takeaway, without diving into the details, is that nearly all models are able to improve and stabilize forecast accuracy once the overfitting threshold is overcome. This is especially true for models which have no built-in feature to treat the overfitting problem. However, even models with such features, like BMA, struggle at first until a certain length of the training set is reached. This confirms our expectation that there is indeed a large variation in results across models and that thus a proper model comparison is needed to gain more insight into the driving factors.

Indeed, there is more to be learned from Figure~\ref{fig:linechart}. The group of (generalized) linear models still contains most competitors (a total of eleven). For the first 35 iterations, the models depict a somewhat similar, erratic forecasting behavior. However, from then onwards, there is a clear separation of the majority of the models towards a MAE region of 0.2, whereas the Bayesian generalized linear model shows an increasingly worse performance. In contrast, (generalized) additive models attain a more stable forecasting pattern earlier in the process. The worst predictions are obtained from exponential smoothing and the Gaussian process, in both we see a large increase in forecasting error at around the 20th iteration. Interestingly, the overall best performance stems from the tree-based models. Specifically, classical random forests and Bayesian additive regression trees are able to accurately pick up the correct signals even in a remarkable overfitting setting, thereby keeping a very stable profile across the iterations. Lastly, the categories of model averaging, support vector machines and neural networks show a somewhat similar pattern of irregularities for the first 30 to 35 iterations and stable performance afterwards. At first glance, it seems that the state-of-the-art credit risk satellite model BMA (red line) already has strong competition.

After descriptively reviewing the evolution of the out-of-sample MAE it would be difficult to state which model performed best and what the actual difference between the models would be. In order to gain more insight into the ranking of the results, we could resort to parametric statistical tests (F-test or t-test) or to their non-parametric alternatives (Friedman test \citep{friedman:1937} or Wilcoxon signed-ranks test \citep{wilcoxon:1945}). However, all of these allow only a pairwise comparison, which would mean 462 comparisons per cross-validation iteration and hence 22,638 in total. Moreover, the parametric tests rely on strong assumptions about the distribution of prediction errors, which seem to be at least partly violated for most models, as can be seen in Figure~\ref{fig:linechart}. Another popular method to compare the accuracy of forecast methods is the pairwise Diebold-Mariano test \citep{diebold1995} and its multivariate alternative \citep{mariano:2012}. However, even with the multivariate test, we would still be left with one test per cross-validation iteration, i.e. 49 tests, if we were to combine all models at once. Even without conducting the test, we can already assume that the equal predictive accuracy (EPA) hypothesis would not hold on such a diverse set of prediction models and thus refrain from implementing the tests.

However, the question of how to compare multiple forecasting models across multiple cross-validation runs has been tackled before. Most notably, \citet{koning:2005} ranked the results of the M3 competition using two non-parametric tests: multiple comparisons with the mean (ANOM) and multiple comparisons with the best (MCB). On the one hand, the new testing strategy allowed them to provide new insights on the statistical significance of the comparative model performance. On the other hand, both tests have received criticism due to the binary nature of the conclusion why certain models performed better or worse. Thus, \citet{demsar:2006} introduced a third test as an alternative based on the Nemenyi test \citep{nemenyi:1963}. The test ranks the performance of each model across the various data sets or cross-validation iterations, averages over the ranks and produces confidence bounds. From the (non-)overlapping confidence bounds one can deduce if the models are statistically different from each other.\footnote{We are aware of the possible drawbacks of using null hypothesis significance testing (see, e.g., \citealp{benavoli:2017} for an overview) and the Bayesian alternatives that could help with such issues (see, e.g. \citealp{calvo:2018}). However, to the best of our knowledge, there is currently no (publicly available) Bayesian hypothesis testing strategy that can handle multiple models and multiple cross-validation results at once.}

\begin{figure}[t!]
    \centering
    \includegraphics[width=1\textwidth]{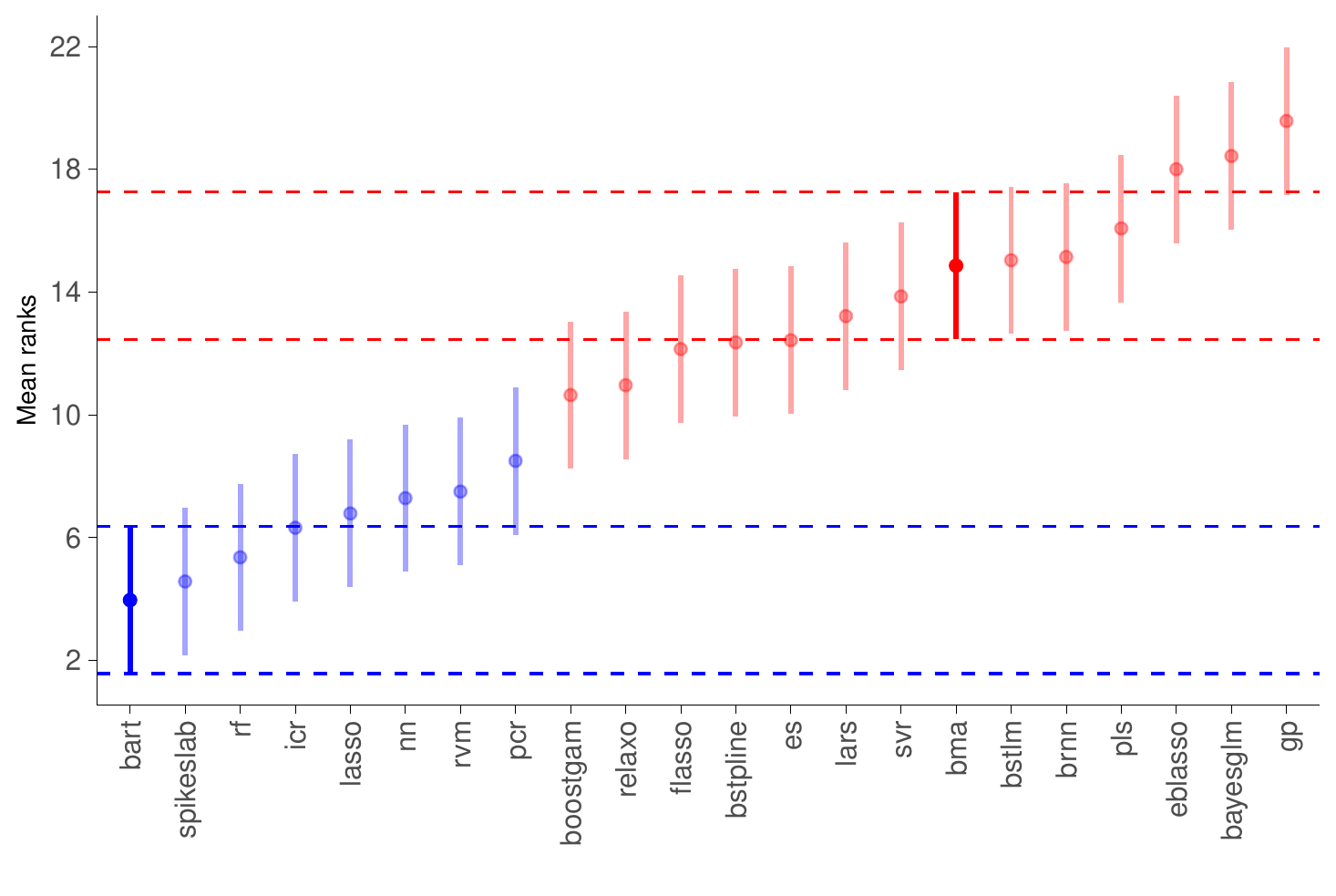}
    \caption{Regression for Multiple Comparison with the Best}
    \fnote{Note: Regression for Multiple Comparison with the Best (RMCB) is the regression-based version of the Nemenyi test \citep{demsar:2006} introduced by \citet{svetunkov:2020}. The models on the x-axis are sorted by the mean rank which they achieved across the cross-validation results, represented by the solid dot. The vertical lines indicate the confidence bounds per model. Bayesian Additive Regression Trees (\textbf{bart}), marked in blue, is the best model. The red indicator depicts the current state-of-the-art model Bayesian Model Averaging (\textbf{bma}). All models that have intersecting confidence bounds with BART or BMA, as shown by the matching colors, are not statistically different from each other. The results are evaluated on a 5\% significance level.}
    \label{fig:nemenyi}
\end{figure}

Figure~\ref{fig:nemenyi} depicts the results of the Regression for Multiple Comparison with the Best (RMCB), which is the regression-based version of the Nemenyi test introduced by \citet{svetunkov:2020}. The test constructs a simple linear regression with the model ranks as dummy variables and uses the estimated coefficients and confidence intervals to determine the performance differences. The main difference between the Nemenyi test and RMCB is the underlying critical distance. For the former it is a atudentized range distribution, while the latter uses a Student's t-distribution. This leads to narrower confidence bounds for the RMCB, which can be helpful with smaller sample sizes.

The models on the x-axis in Figure~\ref{fig:nemenyi} are sorted by the mean rank which they achieved across the cross-validation, depicted by the points in the figure, while the vertical lines reflect the confidence bounds. Thus, this test allows us to reveal the winning model: Bayesian additive regression trees (\textbf{bart}). However, as indicated by the dashed line, there are seven more models that are not statistically different from the winning model on a 5\% significance level. Particularly, BART is closely followed by the spike-and-slab prior (\textbf{spikeslab}) and Random Forests (\textbf{rf}). The following section will provide a deep dive into BART and give more details on the estimation and results.

Within these eight best performing models, two belong to the group of tree-based models (\textbf{bart}, \textbf{rf}), four are (generalized) linear models  (\textbf{spikeslab}, \textbf{icr}, \textbf{lasso}, \textbf{pcr}), one is a neural network (\textbf{nn}) and one is a support/relevance vector machine (\textbf{rvm}). Given the fairly long cross-validation period in which overfitting prevailed (36 out of 49 iterations), it is remarkable that the tree-based models, neural network and relevance vector machine are able to provide such accurate forecasts without the need for regularization. In contrast, it comes as no surprise that all of the linear models use some form of regularization to tackle the overfitting issue.

Lastly, the red indicator depicts again the current state-of-the-art satellite model among central bankers. Within our framework and given these results, we can deduce that Bayesian model averaging (\textbf{bma}) is significantly worse than the first eight models, i.e. the winning group. Hence, for the use case of Austrian corporate default probabilities, these eight models would, on average, deliver more precise forecasts compared to BMA.


\subsection{Deep dive into the winning model}
\label{sec:deepdive}

From the 43 implemented models we started out with and the 22 models that remained, Bayesian additive regression trees (BART) by \citet{chipman:2010} turned out victorious. The following paragraphs will introduce the model in more detail, give insights into why the model worked that well and provide detailed results.

The framework consists of two parts, a sum-of-trees model and regularization priors on the parameters that constraint each tree. BART approximates the function~$f$ from Eq.~\eqref{eq:model} by

\begin{equation}
\label{eq:BART}
    f(\mathbf{X}) \approx \sum_{j=1}^{N} g(\mathbf{X} | \mathcal{T}_{j}, \mathbf{m}_{j}) + \varepsilon, \quad \varepsilon \sim \mathcal{N}(\mathbf{0},\sigma^2\mathbf{I}),
\end{equation}

\noindent whereas $N$ binary trees $\mathcal{T}_{j}$ are used, each $j^{th}$ tree with a vector of $\mathbf{m}_{j} = (\mu_{j1},\dots,\mu_{jb_j})'$ terminal nodes and $b_j$ leaves. 

The second part imposes a set of regularization priors over the grown trees $p(\mathcal{T}_{j})$, the model parameters $p(\mu_{ij}|\mathcal{T}_{j})$ and error variance $p(\sigma)$. These priors ensure that no individual tree is too influential in the sum of trees. Specifically, we want to highlight the prior on the terminal node parameter $\mu_{ij}$, representing the effect of a tree, 

\begin{equation}
\label{eq:prior}
    \mu_{ij} =  N(0,\sigma_{\mu}^2), \quad \sigma_{\mu} = 0.5/k\sqrt{N}.
\end{equation}

\noindent The prior ensures that the tree parameters are shrunken towards zero, thereby constraining each weak learner. The variable $k$ is the prior probability that $E(\mathbf{y}|\mathbf{X})$ is within the range of $\mathbf{y}$. The prior standard deviation $\sigma_{\mu}$ is related to the gradient boosting shrinkage parameter of \citet{friedman:2001}, which also balances the effect of each tree. For more details on the model and inference, we refer the interested reader to the original paper.

The first step in our forecasting framework is the tuning of the hyperparameters. BART, as many other machine learning models, offers the possibility to tune all prior hyperparameters. However, \citet{chipman:2010} also specify out-of-the-box parameters that work well on a range of different data sets. More precisely, one can tune the number of trees $N$, the base ($\alpha$) and power ($\beta$) parameter for the prior $p(\mathcal{T}_{j})$, the above-mentioned variable $k$ for $p(\mu_{ij}|\mathcal{T}_{j})$ and the parameters for the inverse chi-square distribution on $p(\sigma)$, $\nu$ and $q$.

As outlined in Section~\ref{sec:tuning}, we use a grid of potential hyperparameters and the combinations thereof, centered around the default values by \citet{chipman:2010}. Nevertheless, the default values returned the best prediction accuracy, and we will thus not go into the details of the tuning process. There is one exception for which we diverted from the default value, namely the number of trees $N$. \citet{chipman:2010} state a default value of $N=200$ as a larger number of trees increases BART's representation flexibility and thus predictive capabilities. However, they also state that BART can be used for variable selection when the number of trees is reduced. The more trees are grown, the more irrelevant covariates are mixed with relevant ones, diminishing its selection effectiveness. When the number of trees is reduced, BART endogenously picks the more relevant variables. Given our large overfitting period in the training sample, a lower number of tress ($N=50$) achieved the best results.

\begin{figure}[t!]
    \centering
    \includegraphics[width=1\textwidth]{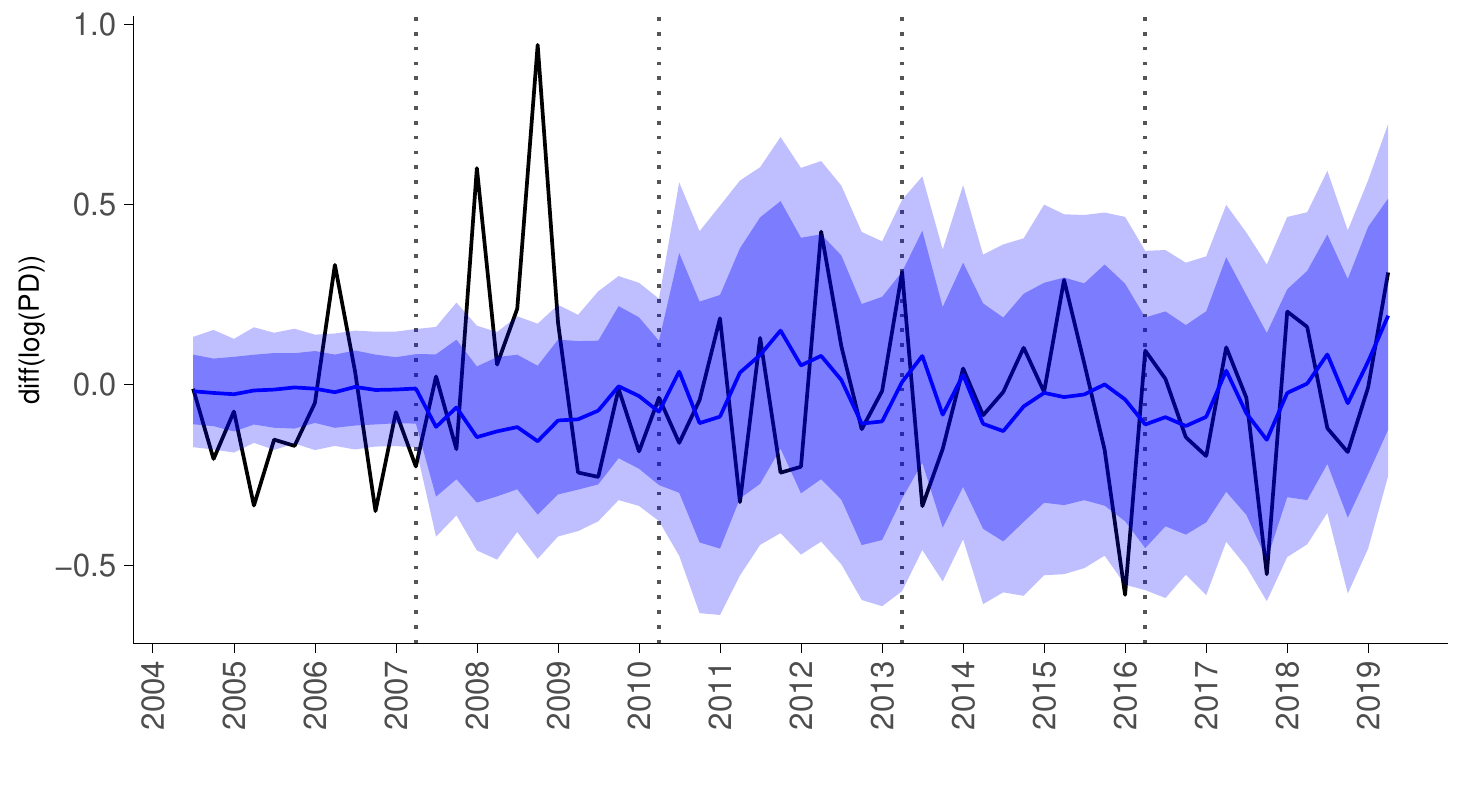}
    \caption{12-step ahead forecasts with prediction intervals}
    \fnote{Note: the figure shows the evolution of the 12-step ahead point prediction and prediction intervals of the BART model. The PD on the y-axis is kept in the same transformation as during the estimation. The solid blue line is the estimated posterior mean, the dark blue shaded area represents the 80\% prediction interval and the light blue area the 95\% interval. The solid black line is the actual PD time series. In order to avoid overlapping lines and ribbons, only five out of 49 cross-validation runs are depicted. The vertical dotted lines indicate the predicted region.}
    \label{fig:predictioninterval}
\end{figure}

Until now, for the purpose of a unified model comparison, we concentrated on point forecasts and the deviation from the true default probability. This was a deliberate choice in order to focus on the prediction accuracy of each model and to keep the set of results manageable. Nevertheless, the uncertainty surrounding the point forecasts is at least as important as the forecast itself. Thus, Figure~\ref{fig:predictioninterval} shows the 12-step ahead forecast from the BART model including prediction intervals. The forecast covers the whole length of the training sample, starting after the initial four quarters as the first cross-validation iteration. The solid blue line is the estimated posterior mean, while the dark blue shaded area represents the 80\% prediction interval and the light blue area the 95\% interval. The solid black line is the actual PD time series. Plotting all cross-validation results (49 in total) would lead to many overlapping points forecasts and indistinguishable prediction intervals. We thus decided to only show five non-overlapping predictions, which nonetheless span all forecasted quarters. The vertical dotted lines indicate the predicted region. Especially in this aspect we can gauge the Bayesian inference as the predictive distribution can simply be calculated from the posterior draws, thereby incorporating the inherent parameter uncertainty.  

Given our long forecasting horizon, there are some deviations between the solid blue and solid black line. Especially in the first two segments -- from around 2004 to 2007 and 2007 to 2010 -- BART was not able to pick up proper signals from the data to detect the surge in default probabilities in the first quarter of 2006 and during the financial crisis of 2007 and 2008. In the subsequent segments, the accuracy of the model increases as the blue line starts to trace the peaks and troughs of the actual PD. The prediction interval keeps a reasonable width over the whole time span. However, the uncertainty around the estimates is undeniable and would need special attention when being used as a credit risk satellite model.

Overall, the non-parametric Bayesian approach coupled with the adaptive weak learners seems to be a very sensitive, capable model that worked well in our setting. Given that many machine learning models depend critically on the chosen set of hyperparameters, the performance of BART with default setting is quite remarkable. The success of BART can be seen in the growing literature in various fields and the technical extensions that have been proposed. Particularly, BART with heteroscedastic errors \citep{pratola:2020}, BART in (non-linear) VAR settings \citep{huber:2020b}, multinomial logistic regression via BART \citep{murray:2017} and survival models \citep{sparapani:2016}. A more detailed overview of the technical extensions can be found in \citet{hill:2020}.


\subsection{Forecast Combinations}
\label{sec:combinations}

As we have seen in last section, even the best models are fraught with uncertainty regarding the point prediction. This inherent uncertainty regarding econometric models is undisputed in the literature and relates to unknown data generating processes, misspecified models and a generally complex reality the models try to replicate. In order to hedge against such uncertainties, \citet{bates:1969} introduced in their seminal paper the concept of forecast combinations. The idea is straight-forward: there is no one true model, there are only different approximations of the data generating process. The underlying models have their own strength and weaknesses, which, when combined, should yield an overall better forecast. Even though such forecast combinations can be hard to interpret in terms of marginal effectiveness of the coefficients, we can again take up the point made by \citet{hastie:2017} that especially with comparisons among prediction models, a stable statistical dependency outweighs the underlying causal relationships.

In the last five decades since \citet{bates:1969}, a wide range of combination methods have been suggested. For the purpose of this paper, we focus on three groups: simple combination methods, regression-based combination methods and eigenvector-based combination methods. For the simple methods, we choose the naive average, which weights all models equally (\textbf{AVG}) and the Newbold/Granger method \citep{newbold:1974}, which calculates the weights from the estimated mean squared prediction error (MSPE) matrix (\textbf{NG}). The second group is still based on linear functions of the individual forecasts, but the weights are determined using a constrained least squares (\textbf{CLS}) regression. Finally, the standard eigenvector-based approach by \citet{hsiao:2014} uses, unlike \citet{newbold:1974}, a normalization condition that leads to an unconstrained minimum of the MSPE (\textbf{SEA}). These methods (and more) have been implemented by \citet{weiss:2018b}.\footnote{Besides a wide array of combination method, the implementation by \citet{weiss:2018b} also allows a dynamic version of combinations, which is related to the idea of time series cross-validation. However, in our setting the normal static version achieved better results than the dynamic version.}

Moreover, we conduct the analysis with three different scenarios: first, in a naive approach we combine all 22 models; second, we only use the winning eight models which have been determined by the Nemenyi test in Figure~\ref{fig:nemenyi}; third, we combine the best model of each category, as outlined in section~\ref{sec:models}.\footnote{Using all 22 models with the Newbold/Granger and the eigenvector-based method leads to mathematical problems and is thus excluded.}

\begin{figure}
    \centering
    \includegraphics[width=1\textwidth]{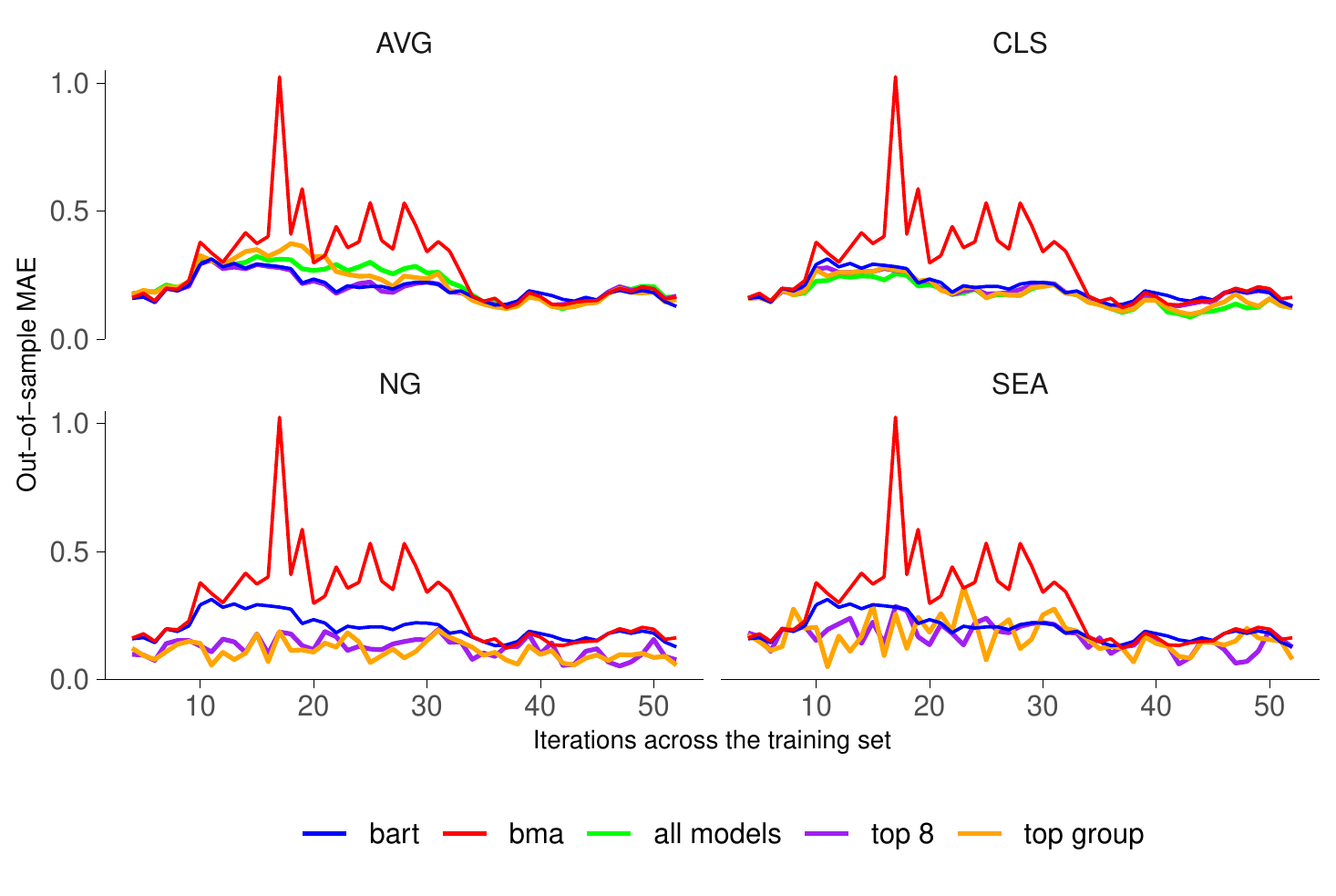}
    \caption{Evolution of out-of-sample MAE for forecast combinations}
    \fnote{Note: The chart depicts the Mean Absolute Error (MAE) for BART (blue line), BMA (red line) and the forecast combination methods: the naive average (AVG), the Newbold/Granger method (NG), the constrained least squares approach (CLS) and the standard eigenvector-based approach (SEA). The forecast combination have been calculated for the scenario using all 22 models (green line), utilizing only the top 8 winning models (purple line) and for the best model per group (orange line).}
    \label{fig:forecastcombination}
\end{figure}

Figure~\ref{fig:forecastcombination} shows, similar to Figure~\ref{fig:linechart}, the evolution of the mean absolute errors (MAE) across the cross-validation iterations. The four panels reflect the forecast combinations group outlined above with the three different model scenarios -- all models (green line), top eight winning models (purple line) and the best models of each category (orange line). In order to properly frame the results of the forecast combinations, we additionally plot the performance of BART (blue line) and BMA (red line). First of all, the sole comparison between BMA and BART again emphasizes the significant increase in terms of predictive accuracy that BART delivers. However, while the simple average method (\textbf{AVG}) is not able to fully outperform BART, the constrained least squares version (\textbf{CLS}) is able to combine the underlying models in a way that beats BART across nearly all cross-validation iterations. The difference becomes even clearer in the case of the Newbold/Granger method (\textbf{NG}) and the standard eigenvector-based approach (\textbf{SEA}). While the eigenvector approach depicts a somewhat erratic behavior, driven by the combined models, the Newbold/Granger approach delivers a significant improvement across all cross-validation iterations.

In order to get a clearer picture of the results, Table~\ref{tab:fc} shows the average out-of-sample MAE and ranks for BART, BMA and the forecast combinations, calculated over the cross-validation iterations. The numbers in parentheses indicate the average forecasting accuracy and rank relative to BART. On average, BART is able to beat two of the three simple average scenarios and is nearly on par with the top eight scenarios. However, all other combination methods are able to improve the average prediction accuracy significantly. The best accuracy is achieved by the Newbold/Granger method using the group-wise best performing models. Although this setup includes rather badly performing models, such as exponential smoothing and the Gaussian process, the Newbold/Granger approach is able to calculate the weights in a way that extract only the positive features from the underlying models.

\begin{table}

    \caption{Average out-of-sample MAE and rank for forecast combinations}
    
    \begin{small}
    
    \centering
    
    \begin{tabular}{ ccccccc }
    \toprule
    & bart & bma & \makecell{AVG \\ all models} & \makecell{AVG \\ top 8} & \makecell{AVG \\ top group} & \makecell{NG \\ top 8} \\
    \cmidrule{2-7}
 
    \multirow{2}{*}{Avg. MAE} & 0.20 & 0.29 & 0.22 & 0.20 & 0.22 & 0.12 \\ 
    & (1.00) & (0.70) & (0.90) & (1.01) & (0.91) & (1.62) \\
        
    \multirow{2}{*}{Avg. Rank} & 8.57 & 10.94 & 10.02 & 8.10 & 8.78 & 2.55 \\ 
    & (1.00) & (0.78) & (0.86) & (1.06) & (0.98) & (3.36) \\ 
     
    & & & & & & \\
     
    & \makecell{NG \\ top group} & \makecell{CLS \\ all models} & \makecell{CLS \\ top 8} & \makecell{CLS \\ top group} & \makecell{SEA \\ top 8} & \makecell{SEA \\ top group} \\
    \cmidrule{2-7}
     
    \multirow{2}{*}{Avg. MAE} & \textbf{0.11} & 0.17 & 0.19 & 0.18 & 0.17 & 0.17 \\ 
    & \textbf{(1.83)} & (1.18) & (1.07) & (1.12) & (1.20) & (1.20) \\
      
    \multirow{2}{*}{Avg. Rank} & \textbf{1.49} & 4.14 & 6.20 & 5.02 & 6.35 & 5.84 \\ 
    & \textbf{(5.75)} & (2.07) & (1.38) & (1.71) & (1.35) & (1.47) \\ 
    \bottomrule

    \end{tabular}
    
    \end{small}
    
    \fnote{Note: The table show the average Mean Absolute Error (MAE) and average rank across the cross-validation results for BART, BMA and the forecast combination methods. The values in the parenthesis are the errors and ranks relative to BART -- values above 1 indicate a better performance.}
    \label{tab:fc}
    
\end{table}


\section{Conclusion}
\label{sec:conclusio}

Since the Great Financial Crisis, the use of stress tests as a tool for assessing the resilience of financial institutions to adverse financial and economic developments has increased significantly. One key part in such exercises is the translation of macroeconomic variables into default probabilities for credit risk by using macrofinancial linkage models. A key requirement for such models is that they should be able to properly detect signals from a wide array of macroeconomic variables in combination with a mostly short data sample. 

The current state-of-the art satellite model for PD translation is Bayesian model averaging (BMA) \citep{raftery:1995}. It has a long track record as being a reliable tool for generating scenario-conditional projections for credit risk and is being adopted by more and more central banks and institutions. However, with the easier access to regression models and the advent of new predictive models in the field of machine learning, the question arises if there are other models that could deliver better results.

The aim of this paper is to conduct a systematic forecast comparison with a large number of different regression models to find the best performing credit risk satellite model. The best model is evaluated for the ability to precisely forecast default probabilities conditional on a standardized set of macroeconomic variables as provided to financial institutions by the \citet{esrb:2020} for the EU-wide banking sector stress test. We implement a total of 43 models which we assigned to 9 categories, ranging from conventional statistical models to more recent machine learning methods. Additionally, we combine the models with different forecast combination approaches to further gauge their potential accuracy. For the purpose of this paper, we implement a framework that allows us to conduct this comparison with a standardized data set for all models, to tune the respective hyperparameters for each model and to cross-validate the results based on recursive pseudo out-of-sample forecasts. The data used in the modelling exercise refer to Austria and include as dependent variable a measure for the probability of default for the non-financial corporate sector and macroeconomic and financial data as independent variables.  

Our results indicate that there are indeed better performing models than the current state-of-the art model. Specifically, a group of eight models significantly outperforms BMA in terms of their capability to forecast default probabilities. Five of these eight models belong to the category of machine learning models. Among these models, there is also the overall winner of the model comparison: Bayesian additive regression trees (BART) by \citet{chipman:2010}. The combination of a flexible sum-of-trees model and well calibrated regularization priors gives BART the advantage needed to outperform all other models, especially in situations where overfitting is prevalent. Additionally, given that the majority of winning models has not been explicitly covered in the literature yet, our comparison sheds light on potential other credit risk models to be further investigated. We specifically highlight the advantages of machine learning models in the context of default probability prediction and more generally their applicability in high dimensions where overfitting prevails. Lastly, as most forecast combinations even outperform BART, we show that simple combination techniques can help to further hedge against model uncertainty and boost predictive accuracy.

\newpage

\begin{landscape}

\appendix
\setcounter{secnumdepth}{0}
\section{Appendix}

\begin{footnotesize}
    \begin{longtable}{ccllll}
      \caption{Overview of implemented models and used R libraries} \\
      \label{table:models} \\
      
      \toprule
      Base Model & Extension & Model Name & Short Name & Libraries & Parameter \\ 
      \midrule
      \endfirsthead
      
         
      \multicolumn{6}{c}{{\bfseries \tablename\ \thetable{} -- continued from previous page}} \\
      \toprule
      Base Model & Extension & Model Name & Short Name & Libraries & Parameter \\ 
      \midrule
      \endhead
         
         
      \multirow{22}{3cm}{\centering (Generalized) Linear Models} 
      
        & Linear & Linear Regression & lm & \texttt{stats} & intercept \\
        \cmidrule{2-6}
        
        & Robust & Robust Linear Model & rlm & \texttt{MASS} & intercept, psi \\
        \cmidrule{2-6} 
        
        & \multirow{8}{*}{Regularized} 
            & \makecell[l]{Linear Regression with\\Forward Selection} & lmfs & \texttt{leaps} & nvmax \\
            \cmidrule{3-6}
            &  & Least Angle Regression & lars & \texttt{lars} & fraction \\
            \cmidrule{3-6}
            &  & Ridge Regression & ridge & \texttt{glmnet} & - \\
            \cmidrule{3-6}
            &  & The lasso & lasso & \texttt{elasticnet} & fraction \\
            \cmidrule{3-6}
            &  & Fused Lasso & flasso & \texttt{penalized} & lambda1, lambda2 \\
            \cmidrule{3-6}
            &  & Relaxed Lasso & relaxo & \texttt{relaxo} & lambda, phi \\
            \cmidrule{3-6}
            &  & Adaptive Lasso & adalasso & \makecell[l]{\texttt{HDeconometrics}, \\ \texttt{glmnet}} & crit \\
            \cmidrule{3-6}
            &  & Penalized GLM & glmnet & \texttt{glmnet} & alpha, lambda \\
            \cmidrule{2-6}
        
        & \multirow{3}{*}{Feature Extraction} 
            & \makecell[l]{Independent Component\\Regression} & icr & \texttt{fastICA} & n.comp \\
            \cmidrule{3-6}
            &  & \makecell[l]{Principal Component\\Analysis} & pca & \texttt{pls} & ncomp \\
            \cmidrule{3-6}
            &  & Partial Least Squares & pls & \texttt{pls} & ncomp \\
            \cmidrule{2-6}
        
        & \multirow{8}{*}{Bayesian} 
           & Bayesian GLM & bglm & \texttt{arm} & - \\
           \cmidrule{3-6}
           &  & Bayesian Ridge Regression & bridge & \texttt{monomvn} & - \\
            \cmidrule{3-6}
            &  & Spike and Slab Regression & spikeslab & \texttt{spikeslab} & vars \\
            \cmidrule{3-6}
            &  & The Bayesian lasso & blasso & \texttt{monomvn} & - \\
            \cmidrule{3-6}
            &  & Empirical Bayesian Lasso  & eblasso & \texttt{eblasso} & a, b \\
            \cmidrule{3-6}
            &  & \makecell[l]{Dirichlet Laplace\\shrinkage prior} & dlbayes & \texttt{dlbayes} & - \\
            \cmidrule{3-6}
            &  & Horseshoe Prior & horseshoe & \texttt{horseshoe} & method.tau \\
            \cmidrule{3-6}
            &  & Horseshoe+ Prior & horseshoe+ & \texttt{bayesreg} & - \\
            \cmidrule{2-6}
        
        & Ensembles & Boosted Linear Model & bstlm & \texttt{bst} & mstop, nu \\
        \cmidrule{1-6}
        
      \multirow{3}{3cm}{\centering Model Averaging} 
      
        & \multirow{2}{*}{Linear}
            & Mallow's Model Averaging & mma & \texttt{mami} & - \\
            \cmidrule{3-6}
            &  & Jackknife Model Averaging & jma & \texttt{mami} & - \\
            \cmidrule{2-6}
        & Bayesian & Bayesian Model Averaging & bma & \texttt{BMS} & - \\
      \cmidrule{1-6}
      
      \multirow{2}{3cm}{\centering Exponential Smoothing} 
      
        & Linear & Exponential Smoothing & es & \texttt{smooth} & ic, xregDo \\
        \cmidrule{2-6} 
        & Complex & \makecell[l]{Complex Exponential\\Smoothing} & ces & \texttt{smooth} & inital, ic, xregDo \\ 
      \cmidrule{1-6}
      
      \multirow{3}{3cm}{\centering (Generalized) Additive Models} 
      
        & Feature Extraction & \makecell[l]{Projection Pursuit\\Regression} & ppr & \texttt{stats} & nterms \\
        \cmidrule{2-6} 
        & Ensembles & \makecell[l]{Boosted Smoothing\\Spline} & bstpline & \texttt{bst} & mstop, nu \\ 
       \cmidrule{2-6}
       & Ensembles & Boosted GAM & boostgam & \texttt{mboost} & mstop, prune \\ 
      \cmidrule{1-6}
      
      \makecell{Multivariate Adaptive\\Regression Splines} & Non-parametric & MARS & mars & \texttt{earth} & nprune, degree \\
      \cmidrule{1-6}
      
      \multirow{2}{3cm}{\centering Support Vector Machines} 
      
        & Regularized & \makecell[l]{SVM with\\Linear Kernel} & svm & \texttt{LiblineaR} & cost, Loss \\
        \cmidrule{2-6}
        & Bayesian & \makecell[l]{RVM with\\Gaussian Kernel} & rvm & \texttt{kernlab} & - \\
      \cmidrule{1-6}
      
      Gaussian Process & Bayesian & \makecell[l]{GP with\\Polynomial Kernel} & gp & \texttt{kernlab} & degree, scale \\
      \cmidrule{1-6}
      
      \multirow{7}{3cm}{\centering Tree-Based Model} 
        & \multirow{2}{*}{Trees} 
           & \makecell[l]{Classification and\\Regression Trees} & cart & \texttt{rpart} & cp \\
           \cmidrule{3-6}
            &  & \makecell[l]{Conditional Inference\\Tree} & ctree & \texttt{party} & \makecell[l]{maxdepth,\\ mincriterion} \\
            \cmidrule{2-6}
            
            & Bayesian & \makecell[l]{Bayesian Additive\\Regression Trees} & bart & \texttt{bartMachine} & \makecell[l]{num\_trees, kvar,\\ alpha, beta, nu} \\
            \cmidrule{2-6}
            
        & \multirow{2}{*}{Ensembles} 
            & Tree-Based Ensembles & enstree & \texttt{nodeHarvest} & maxinter, mode \\
            \cmidrule{3-6}
            & & Gradient Boosted Tree & bsttree & \texttt{bst} & \makecell[l]{mstop,\\maxdepth, nu} \\
            \cmidrule{2-6}
            
        & \multirow{2}{3cm}{\centering Random Forest} 
            & Random Forest & rf & \texttt{ranger}, \texttt{e1071} & mtry, min.node.size\\
            \cmidrule{3-6}
            &  & \makecell[l]{Extremely\\randomized trees} & ert & \texttt{ranger} & \makecell[l]{mtry, splitrule,\\min. node size} \\
      \cmidrule{1-6}
      
      \multirow{2}{3cm}{\centering Neural Network} 
       & Linear & Neural Network & nn & \texttt{neuralnet} & layer1, layer2, layer3 \\
        \cmidrule{2-6}
       & Bayesian & Bayesian NN & brnn & \texttt{brnn} & neurons \\
       
      \bottomrule
      
    \end{longtable} 
\end{footnotesize}

\end{landscape}

\newpage

\newpage

\bibliography{main}

\begin{thebibliography}{}

\bibitem [\protect \citeauthoryear {%
Altman%
\ \BBA {} Bana%
}{%
Altman%
\ \BBA {} Bana%
}{%
{\protect \APACyear {2003}}%
}]{%
altman:2003}
\APACinsertmetastar {%
altman:2003}%
\begin{APACrefauthors}%
Altman, E\BPBI I.%
\BCBT {}\ \BBA {} Bana, G.%
\end{APACrefauthors}%
\unskip\
\newblock
\APACrefYearMonthDay{2003}{}{}.
\newblock
{\BBOQ}\APACrefatitle {Defaults and returns on high yield bonds: The year 2002
  in review and the market outlook} {Defaults and returns on high yield bonds:
  The year 2002 in review and the market outlook}.{\BBCQ}
\newblock

\PrintBackRefs{\CurrentBib}

\bibitem [\protect \citeauthoryear {%
Alves%
}{%
Alves%
}{%
{\protect \APACyear {2005}}%
}]{%
alves:2005}
\APACinsertmetastar {%
alves:2005}%
\begin{APACrefauthors}%
Alves, I.%
\end{APACrefauthors}%
\unskip\
\newblock
\APACrefYearMonthDay{2005}{}{}.
\newblock
{\BBOQ}\APACrefatitle {Sectoral fragility: factors and dynamics} {Sectoral
  fragility: factors and dynamics}.{\BBCQ}
\newblock
\APACjournalVolNumPages{Investigating the relationship between the financial
  and real economy}{22}{}{450--80}.
\PrintBackRefs{\CurrentBib}

\bibitem [\protect \citeauthoryear {%
Armstrong%
}{%
Armstrong%
}{%
{\protect \APACyear {2001}}%
}]{%
armstrong:2001}
\APACinsertmetastar {%
armstrong:2001}%
\begin{APACrefauthors}%
Armstrong, J\BPBI S.%
\end{APACrefauthors}%
\unskip\
\newblock
\APACrefYearMonthDay{2001}{}{}.
\newblock
{\BBOQ}\APACrefatitle {Evaluating Forecasting Methods} {Evaluating forecasting
  methods}.{\BBCQ}
\newblock
\BIn{} J\BPBI S.~Armstrong\ (\BED), \APACrefbtitle {Principles of Forecasting:
  A Handbook for Researchers and Practitioners} {Principles of forecasting: A
  handbook for researchers and practitioners}\ (\BPGS\ 443--472).
\newblock
\APACaddressPublisher{}{Springer}.
\PrintBackRefs{\CurrentBib}

\bibitem [\protect \citeauthoryear {%
Aver%
}{%
Aver%
}{%
{\protect \APACyear {2008}}%
}]{%
aver:2008}
\APACinsertmetastar {%
aver:2008}%
\begin{APACrefauthors}%
Aver, B.%
\end{APACrefauthors}%
\unskip\
\newblock
\APACrefYearMonthDay{2008}{}{}.
\newblock
{\BBOQ}\APACrefatitle {An empirical analysis of credit risk factors of the
  Slovenian banking system} {An empirical analysis of credit risk factors of
  the slovenian banking system}.{\BBCQ}
\newblock
\APACjournalVolNumPages{Managing Global Transitions}{6}{3}{317--334}.
\PrintBackRefs{\CurrentBib}

\bibitem [\protect \citeauthoryear {%
Bates%
\ \BBA {} Granger%
}{%
Bates%
\ \BBA {} Granger%
}{%
{\protect \APACyear {1969}}%
}]{%
bates:1969}
\APACinsertmetastar {%
bates:1969}%
\begin{APACrefauthors}%
Bates, J\BPBI M.%
\BCBT {}\ \BBA {} Granger, C\BPBI W.%
\end{APACrefauthors}%
\unskip\
\newblock
\APACrefYearMonthDay{1969}{}{}.
\newblock
{\BBOQ}\APACrefatitle {The combination of forecasts} {The combination of
  forecasts}.{\BBCQ}
\newblock
\APACjournalVolNumPages{Journal of the Operational Research
  Society}{20}{4}{451--468}.
\PrintBackRefs{\CurrentBib}

\bibitem [\protect \citeauthoryear {%
Benavoli%
, Corani%
, Dem{\v{s}}ar%
\BCBL {}\ \BBA {} Zaffalon%
}{%
Benavoli%
\ \protect \BOthers {.}}{%
{\protect \APACyear {2017}}%
}]{%
benavoli:2017}
\APACinsertmetastar {%
benavoli:2017}%
\begin{APACrefauthors}%
Benavoli, A.%
, Corani, G.%
, Dem{\v{s}}ar, J.%
\BCBL {}\ \BBA {} Zaffalon, M.%
\end{APACrefauthors}%
\unskip\
\newblock
\APACrefYearMonthDay{2017}{}{}.
\newblock
{\BBOQ}\APACrefatitle {{Time for a change: a tutorial for comparing multiple
  classifiers through Bayesian analysis}} {{Time for a change: a tutorial for
  comparing multiple classifiers through Bayesian analysis}}.{\BBCQ}
\newblock
\APACjournalVolNumPages{The Journal of Machine Learning
  Research}{18}{1}{2653--2688}.
\PrintBackRefs{\CurrentBib}

\bibitem [\protect \citeauthoryear {%
Bezanson%
, Edelman%
, Karpinski%
\BCBL {}\ \BBA {} Shah%
}{%
Bezanson%
\ \protect \BOthers {.}}{%
{\protect \APACyear {2017}}%
}]{%
bezanson:2017}
\APACinsertmetastar {%
bezanson:2017}%
\begin{APACrefauthors}%
Bezanson, J.%
, Edelman, A.%
, Karpinski, S.%
\BCBL {}\ \BBA {} Shah, V\BPBI B.%
\end{APACrefauthors}%
\unskip\
\newblock
\APACrefYearMonthDay{2017}{}{}.
\newblock
{\BBOQ}\APACrefatitle {{Julia: A fresh approach to numerical computing}}
  {{Julia: A fresh approach to numerical computing}}.{\BBCQ}
\newblock
\APACjournalVolNumPages{SIAM review}{59}{1}{65--98}.
\PrintBackRefs{\CurrentBib}

\bibitem [\protect \citeauthoryear {%
Bhadra%
, Datta%
, Polson%
\BCBL {}\ \BBA {} Willard%
}{%
Bhadra%
\ \protect \BOthers {.}}{%
{\protect \APACyear {2017}}%
}]{%
bhadra:2017}
\APACinsertmetastar {%
bhadra:2017}%
\begin{APACrefauthors}%
Bhadra, A.%
, Datta, J.%
, Polson, N\BPBI G.%
\BCBL {}\ \BBA {} Willard, B.%
\end{APACrefauthors}%
\unskip\
\newblock
\APACrefYearMonthDay{2017}{}{}.
\newblock
{\BBOQ}\APACrefatitle {The horseshoe+ estimator of ultra-sparse signals} {The
  horseshoe+ estimator of ultra-sparse signals}.{\BBCQ}
\newblock
\APACjournalVolNumPages{Bayesian Analysis}{12}{4}{1105--1131}.
\PrintBackRefs{\CurrentBib}

\bibitem [\protect \citeauthoryear {%
Bhattacharya%
, Pati%
, Pillai%
\BCBL {}\ \BBA {} Dunson%
}{%
Bhattacharya%
\ \protect \BOthers {.}}{%
{\protect \APACyear {2015}}%
}]{%
bhattacharya:2015}
\APACinsertmetastar {%
bhattacharya:2015}%
\begin{APACrefauthors}%
Bhattacharya, A.%
, Pati, D.%
, Pillai, N\BPBI S.%
\BCBL {}\ \BBA {} Dunson, D\BPBI B.%
\end{APACrefauthors}%
\unskip\
\newblock
\APACrefYearMonthDay{2015}{}{}.
\newblock
{\BBOQ}\APACrefatitle {{Dirichlet--Laplace priors for optimal shrinkage}}
  {{Dirichlet--Laplace priors for optimal shrinkage}}.{\BBCQ}
\newblock
\APACjournalVolNumPages{Journal of the American Statistical
  Association}{110}{512}{1479--1490}.
\PrintBackRefs{\CurrentBib}

\bibitem [\protect \citeauthoryear {%
Bofondi%
\ \BBA {} Ropele%
}{%
Bofondi%
\ \BBA {} Ropele%
}{%
{\protect \APACyear {2011}}%
}]{%
bofondi:2011}
\APACinsertmetastar {%
bofondi:2011}%
\begin{APACrefauthors}%
Bofondi, M.%
\BCBT {}\ \BBA {} Ropele, T.%
\end{APACrefauthors}%
\unskip\
\newblock
\APACrefYearMonthDay{2011}{}{}.
\newblock
{\BBOQ}\APACrefatitle {{Macroeconomic determinants of bad loans: evidence from
  Italian banks}} {{Macroeconomic determinants of bad loans: evidence from
  Italian banks}}.{\BBCQ}
\newblock
\APACjournalVolNumPages{Bank of Italy Occasional Paper}{}{89}{}.
\PrintBackRefs{\CurrentBib}

\bibitem [\protect \citeauthoryear {%
Breeden%
}{%
Breeden%
}{%
{\protect \APACyear {2020}}%
}]{%
breeden:2020}
\APACinsertmetastar {%
breeden:2020}%
\begin{APACrefauthors}%
Breeden, J\BPBI L.%
\end{APACrefauthors}%
\unskip\
\newblock
\APACrefYearMonthDay{2020}{}{}.
\newblock
{\BBOQ}\APACrefatitle {A Survey of Machine Learning in Credit Risk} {A survey
  of machine learning in credit risk}.{\BBCQ}
\newblock

\PrintBackRefs{\CurrentBib}

\bibitem [\protect \citeauthoryear {%
Breiman%
}{%
Breiman%
}{%
{\protect \APACyear {2001}}%
}]{%
breiman:2001}
\APACinsertmetastar {%
breiman:2001}%
\begin{APACrefauthors}%
Breiman, L.%
\end{APACrefauthors}%
\unskip\
\newblock
\APACrefYearMonthDay{2001}{}{}.
\newblock
{\BBOQ}\APACrefatitle {Random forests} {Random forests}.{\BBCQ}
\newblock
\APACjournalVolNumPages{Machine learning}{45}{1}{5--32}.
\PrintBackRefs{\CurrentBib}

\bibitem [\protect \citeauthoryear {%
Breiman%
, Friedman%
, Olshen%
\BCBL {}\ \BBA {} Stone%
}{%
Breiman%
\ \protect \BOthers {.}}{%
{\protect \APACyear {1984}}%
}]{%
breiman:1984}
\APACinsertmetastar {%
breiman:1984}%
\begin{APACrefauthors}%
Breiman, L.%
, Friedman, J.%
, Olshen, R.%
\BCBL {}\ \BBA {} Stone, C.%
\end{APACrefauthors}%
\unskip\
\newblock
\APACrefYear{1984}.
\newblock
\APACrefbtitle {Classification and regression trees} {Classification and
  regression trees}.
\newblock
\APACaddressPublisher{}{Wadsworth}.
\PrintBackRefs{\CurrentBib}

\bibitem [\protect \citeauthoryear {%
Brown%
}{%
Brown%
}{%
{\protect \APACyear {1957}}%
}]{%
brown:1957}
\APACinsertmetastar {%
brown:1957}%
\begin{APACrefauthors}%
Brown, R.%
\end{APACrefauthors}%
\unskip\
\newblock
\APACrefYearMonthDay{1957}{}{}.
\newblock
{\BBOQ}\APACrefatitle {Exponential smoothing for predicting demand}
  {Exponential smoothing for predicting demand}.{\BBCQ}
\newblock
\BIn{} \APACrefbtitle {Operations Research} {Operations research}\ (\BVOL~5,
  \BPGS\ 145--145).
\PrintBackRefs{\CurrentBib}

\bibitem [\protect \citeauthoryear {%
Bzdok%
, Altman%
\BCBL {}\ \BBA {} Krzywinski%
}{%
Bzdok%
\ \protect \BOthers {.}}{%
{\protect \APACyear {2018}}%
}]{%
bzdok:2018}
\APACinsertmetastar {%
bzdok:2018}%
\begin{APACrefauthors}%
Bzdok, D.%
, Altman, N.%
\BCBL {}\ \BBA {} Krzywinski, M.%
\end{APACrefauthors}%
\unskip\
\newblock
\APACrefYearMonthDay{2018}{}{}.
\newblock
{\BBOQ}\APACrefatitle {Statistics versus machine learning} {Statistics versus
  machine learning}.{\BBCQ}
\newblock
\APACjournalVolNumPages{Nature methods}{15}{4}{233--234}.
\PrintBackRefs{\CurrentBib}

\bibitem [\protect \citeauthoryear {%
Cai%
, Huang%
\BCBL {}\ \BBA {} Xu%
}{%
Cai%
\ \protect \BOthers {.}}{%
{\protect \APACyear {2011}}%
}]{%
cai:2011}
\APACinsertmetastar {%
cai:2011}%
\begin{APACrefauthors}%
Cai, X.%
, Huang, A.%
\BCBL {}\ \BBA {} Xu, S.%
\end{APACrefauthors}%
\unskip\
\newblock
\APACrefYearMonthDay{2011}{}{}.
\newblock
{\BBOQ}\APACrefatitle {{Fast empirical Bayesian LASSO for multiple quantitative
  trait locus mapping}} {{Fast empirical Bayesian LASSO for multiple
  quantitative trait locus mapping}}.{\BBCQ}
\newblock
\APACjournalVolNumPages{BMC bioinformatics}{12}{1}{1--13}.
\PrintBackRefs{\CurrentBib}

\bibitem [\protect \citeauthoryear {%
Calvo%
, Ceberio%
\BCBL {}\ \BBA {} Lozano%
}{%
Calvo%
\ \protect \BOthers {.}}{%
{\protect \APACyear {2018}}%
}]{%
calvo:2018}
\APACinsertmetastar {%
calvo:2018}%
\begin{APACrefauthors}%
Calvo, B.%
, Ceberio, J.%
\BCBL {}\ \BBA {} Lozano, J\BPBI A.%
\end{APACrefauthors}%
\unskip\
\newblock
\APACrefYearMonthDay{2018}{}{}.
\newblock
{\BBOQ}\APACrefatitle {Bayesian inference for algorithm ranking analysis}
  {Bayesian inference for algorithm ranking analysis}.{\BBCQ}
\newblock
\BIn{} \APACrefbtitle {Proceedings of the Genetic and Evolutionary Computation
  Conference Companion} {Proceedings of the genetic and evolutionary
  computation conference companion}\ (\BPGS\ 324--325).
\PrintBackRefs{\CurrentBib}

\bibitem [\protect \citeauthoryear {%
Carvalho%
, Polson%
\BCBL {}\ \BBA {} Scott%
}{%
Carvalho%
\ \protect \BOthers {.}}{%
{\protect \APACyear {2010}}%
}]{%
carvalho:2010}
\APACinsertmetastar {%
carvalho:2010}%
\begin{APACrefauthors}%
Carvalho, C\BPBI M.%
, Polson, N\BPBI G.%
\BCBL {}\ \BBA {} Scott, J\BPBI G.%
\end{APACrefauthors}%
\unskip\
\newblock
\APACrefYearMonthDay{2010}{}{}.
\newblock
{\BBOQ}\APACrefatitle {The horseshoe estimator for sparse signals} {The
  horseshoe estimator for sparse signals}.{\BBCQ}
\newblock
\APACjournalVolNumPages{Biometrika}{97}{2}{465--480}.
\PrintBackRefs{\CurrentBib}

\bibitem [\protect \citeauthoryear {%
Castren%
, Dees%
\BCBL {}\ \BBA {} Zaher%
}{%
Castren%
\ \protect \BOthers {.}}{%
{\protect \APACyear {2010}}%
}]{%
castren2010}
\APACinsertmetastar {%
castren2010}%
\begin{APACrefauthors}%
Castren, O.%
, Dees, S.%
\BCBL {}\ \BBA {} Zaher, F.%
\end{APACrefauthors}%
\unskip\
\newblock
\APACrefYearMonthDay{2010}{}{}.
\newblock
{\BBOQ}\APACrefatitle {{Stress-testing euro area corporate default
  probabilities using a global macroeconomic model}} {{Stress-testing euro area
  corporate default probabilities using a global macroeconomic model}}.{\BBCQ}
\newblock
\APACjournalVolNumPages{Journal of Financial Stability}{6}{2}{64--78}.
\PrintBackRefs{\CurrentBib}

\bibitem [\protect \citeauthoryear {%
Castro%
}{%
Castro%
}{%
{\protect \APACyear {2013}}%
}]{%
castro:2013}
\APACinsertmetastar {%
castro:2013}%
\begin{APACrefauthors}%
Castro, V.%
\end{APACrefauthors}%
\unskip\
\newblock
\APACrefYearMonthDay{2013}{}{}.
\newblock
{\BBOQ}\APACrefatitle {{Macroeconomic determinants of the credit risk in the
  banking system: The case of the GIPSI}} {{Macroeconomic determinants of the
  credit risk in the banking system: The case of the GIPSI}}.{\BBCQ}
\newblock
\APACjournalVolNumPages{Economic Modelling}{31}{}{672--683}.
\PrintBackRefs{\CurrentBib}

\bibitem [\protect \citeauthoryear {%
Chipman%
, George%
\BCBL {}\ \BBA {} McCulloch%
}{%
Chipman%
\ \protect \BOthers {.}}{%
{\protect \APACyear {2010}}%
}]{%
chipman:2010}
\APACinsertmetastar {%
chipman:2010}%
\begin{APACrefauthors}%
Chipman, H\BPBI A.%
, George, E\BPBI I.%
\BCBL {}\ \BBA {} McCulloch, R\BPBI E.%
\end{APACrefauthors}%
\unskip\
\newblock
\APACrefYearMonthDay{2010}{}{}.
\newblock
{\BBOQ}\APACrefatitle {BART: Bayesian additive regression trees} {Bart:
  Bayesian additive regression trees}.{\BBCQ}
\newblock
\APACjournalVolNumPages{The Annals of Applied Statistics}{4}{1}{266--298}.
\PrintBackRefs{\CurrentBib}

\bibitem [\protect \citeauthoryear {%
Comon%
}{%
Comon%
}{%
{\protect \APACyear {1994}}%
}]{%
comon:1994}
\APACinsertmetastar {%
comon:1994}%
\begin{APACrefauthors}%
Comon, P.%
\end{APACrefauthors}%
\unskip\
\newblock
\APACrefYearMonthDay{1994}{}{}.
\newblock
{\BBOQ}\APACrefatitle {Independent component analysis, a new concept?}
  {Independent component analysis, a new concept?}{\BBCQ}
\newblock
\APACjournalVolNumPages{Signal processing}{36}{3}{287--314}.
\PrintBackRefs{\CurrentBib}

\bibitem [\protect \citeauthoryear {%
Dem{\v{s}}ar%
}{%
Dem{\v{s}}ar%
}{%
{\protect \APACyear {2006}}%
}]{%
demsar:2006}
\APACinsertmetastar {%
demsar:2006}%
\begin{APACrefauthors}%
Dem{\v{s}}ar, J.%
\end{APACrefauthors}%
\unskip\
\newblock
\APACrefYearMonthDay{2006}{}{}.
\newblock
{\BBOQ}\APACrefatitle {Statistical comparisons of classifiers over multiple
  data sets} {Statistical comparisons of classifiers over multiple data
  sets}.{\BBCQ}
\newblock
\APACjournalVolNumPages{The Journal of Machine Learning Research}{7}{}{1--30}.
\PrintBackRefs{\CurrentBib}

\bibitem [\protect \citeauthoryear {%
Dickey%
\ \BBA {} Fuller%
}{%
Dickey%
\ \BBA {} Fuller%
}{%
{\protect \APACyear {1979}}%
}]{%
dickey:1979}
\APACinsertmetastar {%
dickey:1979}%
\begin{APACrefauthors}%
Dickey, D\BPBI A.%
\BCBT {}\ \BBA {} Fuller, W\BPBI A.%
\end{APACrefauthors}%
\unskip\
\newblock
\APACrefYearMonthDay{1979}{}{}.
\newblock
{\BBOQ}\APACrefatitle {Distribution of the estimators for autoregressive time
  series with a unit root} {Distribution of the estimators for autoregressive
  time series with a unit root}.{\BBCQ}
\newblock
\APACjournalVolNumPages{Journal of the American statistical
  association}{74}{366a}{427--431}.
\PrintBackRefs{\CurrentBib}

\bibitem [\protect \citeauthoryear {%
Diebold%
\ \BBA {} Mariano%
}{%
Diebold%
\ \BBA {} Mariano%
}{%
{\protect \APACyear {1995}}%
}]{%
diebold1995}
\APACinsertmetastar {%
diebold1995}%
\begin{APACrefauthors}%
Diebold, F\BPBI X.%
\BCBT {}\ \BBA {} Mariano, R\BPBI S.%
\end{APACrefauthors}%
\unskip\
\newblock
\APACrefYearMonthDay{1995}{}{}.
\newblock
{\BBOQ}\APACrefatitle {Comparing predictive accuracy} {Comparing predictive
  accuracy}.{\BBCQ}
\newblock
\APACjournalVolNumPages{Journal of Business and Economic
  Statistics}{13}{3}{253--263}.
\PrintBackRefs{\CurrentBib}

\bibitem [\protect \citeauthoryear {%
Drucker%
, Burges%
, Kaufman%
, Smola%
\BCBL {}\ \BBA {} Vapnik%
}{%
Drucker%
\ \protect \BOthers {.}}{%
{\protect \APACyear {1997}}%
}]{%
drucker:1997}
\APACinsertmetastar {%
drucker:1997}%
\begin{APACrefauthors}%
Drucker, H.%
, Burges, C.%
, Kaufman, L.%
, Smola, A.%
\BCBL {}\ \BBA {} Vapnik, V.%
\end{APACrefauthors}%
\unskip\
\newblock
\APACrefYearMonthDay{1997}{}{}.
\newblock
{\BBOQ}\APACrefatitle {Support vector regression machines} {Support vector
  regression machines}.{\BBCQ}
\newblock
\APACjournalVolNumPages{Advances in neural information processing
  systems}{9}{}{155--161}.
\PrintBackRefs{\CurrentBib}

\bibitem [\protect \citeauthoryear {%
ECB%
}{%
ECB%
}{%
{\protect \APACyear {2020}}%
}]{%
ecb:HICP}
\APACinsertmetastar {%
ecb:HICP}%
\begin{APACrefauthors}%
ECB.%
\end{APACrefauthors}%
\unskip\
\newblock
\APACrefYearMonthDay{2020}{}{}.
\newblock
\APACrefbtitle {{The definition of price stability}.} {{The definition of price
  stability}.}
\newblock
\begin{APACrefURL}
  [{2020-11-06}]\url{https://www.ecb.europa.eu/mopo/strategy/pricestab/html/index.en.html}
  \end{APACrefURL}
\PrintBackRefs{\CurrentBib}

\bibitem [\protect \citeauthoryear {%
Efron%
, Hastie%
, Johnstone%
\BCBL {}\ \BBA {} Tibshirani%
}{%
Efron%
\ \protect \BOthers {.}}{%
{\protect \APACyear {2004}}%
}]{%
efron:2004}
\APACinsertmetastar {%
efron:2004}%
\begin{APACrefauthors}%
Efron, B.%
, Hastie, T.%
, Johnstone, I.%
\BCBL {}\ \BBA {} Tibshirani, R.%
\end{APACrefauthors}%
\unskip\
\newblock
\APACrefYearMonthDay{2004}{}{}.
\newblock
{\BBOQ}\APACrefatitle {Least angle regression} {Least angle regression}.{\BBCQ}
\newblock
\APACjournalVolNumPages{Annals of statistics}{32}{2}{407--499}.
\PrintBackRefs{\CurrentBib}

\bibitem [\protect \citeauthoryear {%
Elliott%
, Rothenberg%
\BCBL {}\ \BBA {} Stock%
}{%
Elliott%
\ \protect \BOthers {.}}{%
{\protect \APACyear {1996}}%
}]{%
elliott:1996}
\APACinsertmetastar {%
elliott:1996}%
\begin{APACrefauthors}%
Elliott, G.%
, Rothenberg, T\BPBI J.%
\BCBL {}\ \BBA {} Stock, J\BPBI H.%
\end{APACrefauthors}%
\unskip\
\newblock
\APACrefYearMonthDay{1996}{}{}.
\newblock
{\BBOQ}\APACrefatitle {{Efficient Tests for an Autoregressive Unit Root}}
  {{Efficient Tests for an Autoregressive Unit Root}}.{\BBCQ}
\newblock
\APACjournalVolNumPages{Econometrica}{64}{4}{813–-836}.
\PrintBackRefs{\CurrentBib}

\bibitem [\protect \citeauthoryear {%
ESRB%
}{%
ESRB%
}{%
{\protect \APACyear {2020}}%
}]{%
esrb:2020}
\APACinsertmetastar {%
esrb:2020}%
\begin{APACrefauthors}%
ESRB.%
\end{APACrefauthors}%
\unskip\
\newblock
\APACrefYearMonthDay{2020}{}{}.
\newblock
\APACrefbtitle {{Macro-financial scenario for the 2020 EU-wide banking sector
  stress test}.} {{Macro-financial scenario for the 2020 EU-wide banking sector
  stress test}.}
\newblock
\begin{APACrefURL}
  [{2020-11-06}]\url{https://www.esrb.europa.eu/mppa/stress/shared/pdf/esrb.stress_test200131~09dbe748d4.en.pdf}
  \end{APACrefURL}
\PrintBackRefs{\CurrentBib}

\bibitem [\protect \citeauthoryear {%
Fan%
\ \BBA {} Li%
}{%
Fan%
\ \BBA {} Li%
}{%
{\protect \APACyear {2001}}%
}]{%
fan:2001}
\APACinsertmetastar {%
fan:2001}%
\begin{APACrefauthors}%
Fan, J.%
\BCBT {}\ \BBA {} Li, R.%
\end{APACrefauthors}%
\unskip\
\newblock
\APACrefYearMonthDay{2001}{}{}.
\newblock
{\BBOQ}\APACrefatitle {Variable selection via nonconcave penalized likelihood
  and its oracle properties} {Variable selection via nonconcave penalized
  likelihood and its oracle properties}.{\BBCQ}
\newblock
\APACjournalVolNumPages{Journal of the American statistical
  Association}{96}{456}{1348--1360}.
\PrintBackRefs{\CurrentBib}

\bibitem [\protect \citeauthoryear {%
Feurer%
\ \BBA {} Hutter%
}{%
Feurer%
\ \BBA {} Hutter%
}{%
{\protect \APACyear {2019}}%
}]{%
feurer:2019}
\APACinsertmetastar {%
feurer:2019}%
\begin{APACrefauthors}%
Feurer, M.%
\BCBT {}\ \BBA {} Hutter, F.%
\end{APACrefauthors}%
\unskip\
\newblock
\APACrefYearMonthDay{2019}{}{}.
\newblock
{\BBOQ}\APACrefatitle {{Hyperparameter Optimization}} {{Hyperparameter
  Optimization}}.{\BBCQ}
\newblock
\BIn{} \APACrefbtitle {Automated machine learning: methods, systems,
  challenges} {Automated machine learning: methods, systems, challenges}\
  (\BPGS\ 3--33).
\PrintBackRefs{\CurrentBib}

\bibitem [\protect \citeauthoryear {%
Foresee%
\ \BBA {} Hagan%
}{%
Foresee%
\ \BBA {} Hagan%
}{%
{\protect \APACyear {1997}}%
}]{%
foresee:1997}
\APACinsertmetastar {%
foresee:1997}%
\begin{APACrefauthors}%
Foresee, D.%
\BCBT {}\ \BBA {} Hagan, M.%
\end{APACrefauthors}%
\unskip\
\newblock
\APACrefYearMonthDay{1997}{}{}.
\newblock
{\BBOQ}\APACrefatitle {Gauss-Newton approximation to Bayesian learning}
  {Gauss-newton approximation to bayesian learning}.{\BBCQ}
\newblock
\BIn{} \APACrefbtitle {Proceedings of international conference on neural
  networks (ICNN'97)} {Proceedings of international conference on neural
  networks (icnn'97)}\ (\BVOL~3, \BPGS\ 1930--1935).
\PrintBackRefs{\CurrentBib}

\bibitem [\protect \citeauthoryear {%
J.~Friedman%
}{%
J.~Friedman%
}{%
{\protect \APACyear {1991}}%
}]{%
friedman:1991}
\APACinsertmetastar {%
friedman:1991}%
\begin{APACrefauthors}%
Friedman, J.%
\end{APACrefauthors}%
\unskip\
\newblock
\APACrefYearMonthDay{1991}{}{}.
\newblock
{\BBOQ}\APACrefatitle {Multivariate adaptive regression splines} {Multivariate
  adaptive regression splines}.{\BBCQ}
\newblock
\APACjournalVolNumPages{The annals of statistics}{}{}{1--67}.
\PrintBackRefs{\CurrentBib}

\bibitem [\protect \citeauthoryear {%
J.~Friedman%
}{%
J.~Friedman%
}{%
{\protect \APACyear {2001}}%
}]{%
friedman:2001}
\APACinsertmetastar {%
friedman:2001}%
\begin{APACrefauthors}%
Friedman, J.%
\end{APACrefauthors}%
\unskip\
\newblock
\APACrefYearMonthDay{2001}{}{}.
\newblock
{\BBOQ}\APACrefatitle {Greedy function approximation: a gradient boosting
  machine} {Greedy function approximation: a gradient boosting machine}.{\BBCQ}
\newblock
\APACjournalVolNumPages{Annals of statistics}{}{}{1189--1232}.
\PrintBackRefs{\CurrentBib}

\bibitem [\protect \citeauthoryear {%
J.~Friedman%
\ \BBA {} Stuetzle%
}{%
J.~Friedman%
\ \BBA {} Stuetzle%
}{%
{\protect \APACyear {1981}}%
}]{%
friedman:1981}
\APACinsertmetastar {%
friedman:1981}%
\begin{APACrefauthors}%
Friedman, J.%
\BCBT {}\ \BBA {} Stuetzle, W.%
\end{APACrefauthors}%
\unskip\
\newblock
\APACrefYearMonthDay{1981}{}{}.
\newblock
{\BBOQ}\APACrefatitle {Projection pursuit regression} {Projection pursuit
  regression}.{\BBCQ}
\newblock
\APACjournalVolNumPages{Journal of the American statistical
  Association}{76}{376}{817--823}.
\PrintBackRefs{\CurrentBib}

\bibitem [\protect \citeauthoryear {%
M.~Friedman%
}{%
M.~Friedman%
}{%
{\protect \APACyear {1937}}%
}]{%
friedman:1937}
\APACinsertmetastar {%
friedman:1937}%
\begin{APACrefauthors}%
Friedman, M.%
\end{APACrefauthors}%
\unskip\
\newblock
\APACrefYearMonthDay{1937}{}{}.
\newblock
{\BBOQ}\APACrefatitle {The use of ranks to avoid the assumption of normality
  implicit in the analysis of variance} {The use of ranks to avoid the
  assumption of normality implicit in the analysis of variance}.{\BBCQ}
\newblock
\APACjournalVolNumPages{Journal of the American statistical
  association}{32}{200}{675--701}.
\PrintBackRefs{\CurrentBib}

\bibitem [\protect \citeauthoryear {%
Gambera%
}{%
Gambera%
}{%
{\protect \APACyear {2000}}%
}]{%
gambera:2000}
\APACinsertmetastar {%
gambera:2000}%
\begin{APACrefauthors}%
Gambera, M.%
\end{APACrefauthors}%
\unskip\
\newblock
\APACrefYear{2000}.
\newblock
\APACrefbtitle {Simple forecasts of bank loan quality in the business cycle}
  {Simple forecasts of bank loan quality in the business cycle}\ (\BVOL~230).
\newblock
\APACaddressPublisher{}{Federal Reserve Bank of Chicago Chicago, IL}.
\PrintBackRefs{\CurrentBib}

\bibitem [\protect \citeauthoryear {%
Gelman%
, Jakulin%
, Pittau%
\BCBL {}\ \BBA {} Su%
}{%
Gelman%
\ \protect \BOthers {.}}{%
{\protect \APACyear {2008}}%
}]{%
gelman:2008}
\APACinsertmetastar {%
gelman:2008}%
\begin{APACrefauthors}%
Gelman, A.%
, Jakulin, A.%
, Pittau, M\BPBI G.%
\BCBL {}\ \BBA {} Su, Y\BHBI S.%
\end{APACrefauthors}%
\unskip\
\newblock
\APACrefYearMonthDay{2008}{}{}.
\newblock
{\BBOQ}\APACrefatitle {A weakly informative default prior distribution for
  logistic and other regression models} {A weakly informative default prior
  distribution for logistic and other regression models}.{\BBCQ}
\newblock
\APACjournalVolNumPages{Annals of applied Statistics}{2}{4}{1360--1383}.
\PrintBackRefs{\CurrentBib}

\bibitem [\protect \citeauthoryear {%
Geurts%
, Ernst%
\BCBL {}\ \BBA {} Wehenkel%
}{%
Geurts%
\ \protect \BOthers {.}}{%
{\protect \APACyear {2006}}%
}]{%
geurts:2006}
\APACinsertmetastar {%
geurts:2006}%
\begin{APACrefauthors}%
Geurts, P.%
, Ernst, D.%
\BCBL {}\ \BBA {} Wehenkel, L.%
\end{APACrefauthors}%
\unskip\
\newblock
\APACrefYearMonthDay{2006}{}{}.
\newblock
{\BBOQ}\APACrefatitle {Extremely randomized trees} {Extremely randomized
  trees}.{\BBCQ}
\newblock
\APACjournalVolNumPages{Machine learning}{63}{1}{3--42}.
\PrintBackRefs{\CurrentBib}

\bibitem [\protect \citeauthoryear {%
Gross%
\ \BBA {} Poblaci{\'o}n%
}{%
Gross%
\ \BBA {} Poblaci{\'o}n%
}{%
{\protect \APACyear {2019}}%
}]{%
gross:2019}
\APACinsertmetastar {%
gross:2019}%
\begin{APACrefauthors}%
Gross, M.%
\BCBT {}\ \BBA {} Poblaci{\'o}n, J.%
\end{APACrefauthors}%
\unskip\
\newblock
\APACrefYearMonthDay{2019}{}{}.
\newblock
{\BBOQ}\APACrefatitle {{Implications of model uncertainty for bank stress
  testing}} {{Implications of model uncertainty for bank stress
  testing}}.{\BBCQ}
\newblock
\APACjournalVolNumPages{Journal of Financial Services Research}{55}{1}{31--58}.
\PrintBackRefs{\CurrentBib}

\bibitem [\protect \citeauthoryear {%
Grundke%
, Pliszka%
\BCBL {}\ \BBA {} Tuchscherer%
}{%
Grundke%
\ \protect \BOthers {.}}{%
{\protect \APACyear {2019}}%
}]{%
grundke:2019}
\APACinsertmetastar {%
grundke:2019}%
\begin{APACrefauthors}%
Grundke, P.%
, Pliszka, K.%
\BCBL {}\ \BBA {} Tuchscherer, M.%
\end{APACrefauthors}%
\unskip\
\newblock
\APACrefYearMonthDay{2019}{}{}.
\newblock
{\BBOQ}\APACrefatitle {Model and estimation risk in credit risk stress tests}
  {Model and estimation risk in credit risk stress tests}.{\BBCQ}
\newblock
\APACjournalVolNumPages{Review of Quantitative Finance and
  Accounting}{}{}{1--37}.
\PrintBackRefs{\CurrentBib}

\bibitem [\protect \citeauthoryear {%
Hansen%
}{%
Hansen%
}{%
{\protect \APACyear {2007}}%
}]{%
hansen:2007}
\APACinsertmetastar {%
hansen:2007}%
\begin{APACrefauthors}%
Hansen, B.%
\end{APACrefauthors}%
\unskip\
\newblock
\APACrefYearMonthDay{2007}{}{}.
\newblock
{\BBOQ}\APACrefatitle {Least squares model averaging} {Least squares model
  averaging}.{\BBCQ}
\newblock
\APACjournalVolNumPages{Econometrica}{75}{4}{1175--1189}.
\PrintBackRefs{\CurrentBib}

\bibitem [\protect \citeauthoryear {%
Hansen%
\ \BBA {} Racine%
}{%
Hansen%
\ \BBA {} Racine%
}{%
{\protect \APACyear {2012}}%
}]{%
hansen:2012}
\APACinsertmetastar {%
hansen:2012}%
\begin{APACrefauthors}%
Hansen, B.%
\BCBT {}\ \BBA {} Racine, J.%
\end{APACrefauthors}%
\unskip\
\newblock
\APACrefYearMonthDay{2012}{}{}.
\newblock
{\BBOQ}\APACrefatitle {Jackknife model averaging} {Jackknife model
  averaging}.{\BBCQ}
\newblock
\APACjournalVolNumPages{Journal of Econometrics}{167}{1}{38--46}.
\PrintBackRefs{\CurrentBib}

\bibitem [\protect \citeauthoryear {%
Hastie%
, Tibshirani%
\BCBL {}\ \BBA {} Tibshirani%
}{%
Hastie%
\ \protect \BOthers {.}}{%
{\protect \APACyear {2017}}%
}]{%
hastie:2017}
\APACinsertmetastar {%
hastie:2017}%
\begin{APACrefauthors}%
Hastie, T.%
, Tibshirani, R.%
\BCBL {}\ \BBA {} Tibshirani, R\BPBI J.%
\end{APACrefauthors}%
\unskip\
\newblock
\APACrefYearMonthDay{2017}{}{}.
\newblock
{\BBOQ}\APACrefatitle {Extended comparisons of best subset selection, forward
  stepwise selection, and the lasso} {Extended comparisons of best subset
  selection, forward stepwise selection, and the lasso}.{\BBCQ}
\newblock
\APACjournalVolNumPages{arXiv preprint arXiv:1707.08692}{}{}{}.
\PrintBackRefs{\CurrentBib}

\bibitem [\protect \citeauthoryear {%
Hill%
, Linero%
\BCBL {}\ \BBA {} Murray%
}{%
Hill%
\ \protect \BOthers {.}}{%
{\protect \APACyear {2020}}%
}]{%
hill:2020}
\APACinsertmetastar {%
hill:2020}%
\begin{APACrefauthors}%
Hill, J.%
, Linero, A.%
\BCBL {}\ \BBA {} Murray, J.%
\end{APACrefauthors}%
\unskip\
\newblock
\APACrefYearMonthDay{2020}{}{}.
\newblock
{\BBOQ}\APACrefatitle {Bayesian additive regression trees: a review and look
  forward} {Bayesian additive regression trees: a review and look
  forward}.{\BBCQ}
\newblock
\APACjournalVolNumPages{Annual Review of Statistics and Its
  Application}{7}{}{251--278}.
\PrintBackRefs{\CurrentBib}

\bibitem [\protect \citeauthoryear {%
Hocking%
}{%
Hocking%
}{%
{\protect \APACyear {1976}}%
}]{%
hocking:1976}
\APACinsertmetastar {%
hocking:1976}%
\begin{APACrefauthors}%
Hocking, R\BPBI R.%
\end{APACrefauthors}%
\unskip\
\newblock
\APACrefYearMonthDay{1976}{}{}.
\newblock
{\BBOQ}\APACrefatitle {The analysis and selection of variables in linear
  regression} {The analysis and selection of variables in linear
  regression}.{\BBCQ}
\newblock
\APACjournalVolNumPages{Biometrics}{}{}{1--49}.
\PrintBackRefs{\CurrentBib}

\bibitem [\protect \citeauthoryear {%
Hothorn%
, Hornik%
\BCBL {}\ \BBA {} Zeileis%
}{%
Hothorn%
\ \protect \BOthers {.}}{%
{\protect \APACyear {2006}}%
}]{%
hothorn:2006}
\APACinsertmetastar {%
hothorn:2006}%
\begin{APACrefauthors}%
Hothorn, T.%
, Hornik, K.%
\BCBL {}\ \BBA {} Zeileis, A.%
\end{APACrefauthors}%
\unskip\
\newblock
\APACrefYearMonthDay{2006}{}{}.
\newblock
{\BBOQ}\APACrefatitle {Unbiased recursive partitioning: A conditional inference
  framework} {Unbiased recursive partitioning: A conditional inference
  framework}.{\BBCQ}
\newblock
\APACjournalVolNumPages{Journal of Computational and Graphical
  statistics}{15}{3}{651--674}.
\PrintBackRefs{\CurrentBib}

\bibitem [\protect \citeauthoryear {%
Hsiao%
\ \BBA {} Wan%
}{%
Hsiao%
\ \BBA {} Wan%
}{%
{\protect \APACyear {2014}}%
}]{%
hsiao:2014}
\APACinsertmetastar {%
hsiao:2014}%
\begin{APACrefauthors}%
Hsiao, C.%
\BCBT {}\ \BBA {} Wan, S\BPBI K.%
\end{APACrefauthors}%
\unskip\
\newblock
\APACrefYearMonthDay{2014}{}{}.
\newblock
{\BBOQ}\APACrefatitle {Is there an optimal forecast combination?} {Is there an
  optimal forecast combination?}{\BBCQ}
\newblock
\APACjournalVolNumPages{Journal of Econometrics}{178}{}{294--309}.
\PrintBackRefs{\CurrentBib}

\bibitem [\protect \citeauthoryear {%
Hsieh%
, Chang%
, Lin%
, Keerthi%
\BCBL {}\ \BBA {} Sundararajan%
}{%
Hsieh%
\ \protect \BOthers {.}}{%
{\protect \APACyear {2008}}%
}]{%
hsieh:2008}
\APACinsertmetastar {%
hsieh:2008}%
\begin{APACrefauthors}%
Hsieh, C\BHBI J.%
, Chang, K\BHBI W.%
, Lin, C\BHBI J.%
, Keerthi, S.%
\BCBL {}\ \BBA {} Sundararajan, S.%
\end{APACrefauthors}%
\unskip\
\newblock
\APACrefYearMonthDay{2008}{}{}.
\newblock
{\BBOQ}\APACrefatitle {A dual coordinate descent method for large-scale linear
  SVM} {A dual coordinate descent method for large-scale linear svm}.{\BBCQ}
\newblock
\BIn{} \APACrefbtitle {Proceedings of the 25th international conference on
  Machine learning} {Proceedings of the 25th international conference on
  machine learning}\ (\BPGS\ 408--415).
\PrintBackRefs{\CurrentBib}

\bibitem [\protect \citeauthoryear {%
F.~Huber%
, Koop%
, Onorante%
, Pfarrhofer%
\BCBL {}\ \BBA {} Schreiner%
}{%
F.~Huber%
\ \protect \BOthers {.}}{%
{\protect \APACyear {2020}}%
}]{%
huber:2020b}
\APACinsertmetastar {%
huber:2020b}%
\begin{APACrefauthors}%
Huber, F.%
, Koop, G.%
, Onorante, L.%
, Pfarrhofer, M.%
\BCBL {}\ \BBA {} Schreiner, J.%
\end{APACrefauthors}%
\unskip\
\newblock
\APACrefYearMonthDay{2020}{}{}.
\newblock
{\BBOQ}\APACrefatitle {Nowcasting in a pandemic using non-parametric mixed
  frequency VARs} {Nowcasting in a pandemic using non-parametric mixed
  frequency vars}.{\BBCQ}
\newblock
\APACjournalVolNumPages{Journal of Econometrics}{}{}{}.
\PrintBackRefs{\CurrentBib}

\bibitem [\protect \citeauthoryear {%
P.~Huber%
}{%
P.~Huber%
}{%
{\protect \APACyear {1992}}%
}]{%
huber:1992}
\APACinsertmetastar {%
huber:1992}%
\begin{APACrefauthors}%
Huber, P.%
\end{APACrefauthors}%
\unskip\
\newblock
\APACrefYearMonthDay{1992}{}{}.
\newblock
{\BBOQ}\APACrefatitle {Robust estimation of a location parameter} {Robust
  estimation of a location parameter}.{\BBCQ}
\newblock
\BIn{} \APACrefbtitle {Breakthroughs in statistics} {Breakthroughs in
  statistics}\ (\BPGS\ 492--518).
\newblock
\APACaddressPublisher{}{Springer}.
\PrintBackRefs{\CurrentBib}

\bibitem [\protect \citeauthoryear {%
Hyndman%
\ \BBA {} Koehler%
}{%
Hyndman%
\ \BBA {} Koehler%
}{%
{\protect \APACyear {2006}}%
}]{%
hyndman:2006}
\APACinsertmetastar {%
hyndman:2006}%
\begin{APACrefauthors}%
Hyndman, R.%
\BCBT {}\ \BBA {} Koehler, A.%
\end{APACrefauthors}%
\unskip\
\newblock
\APACrefYearMonthDay{2006}{}{}.
\newblock
{\BBOQ}\APACrefatitle {Another look at measures of forecast accuracy} {Another
  look at measures of forecast accuracy}.{\BBCQ}
\newblock
\APACjournalVolNumPages{International journal of forecasting}{22}{4}{679--688}.
\PrintBackRefs{\CurrentBib}

\bibitem [\protect \citeauthoryear {%
Hyndman%
, Koehler%
, Ord%
\BCBL {}\ \BBA {} Snyder%
}{%
Hyndman%
\ \protect \BOthers {.}}{%
{\protect \APACyear {2008}}%
}]{%
hyndman:2008}
\APACinsertmetastar {%
hyndman:2008}%
\begin{APACrefauthors}%
Hyndman, R.%
, Koehler, A.%
, Ord, K.%
\BCBL {}\ \BBA {} Snyder, R.%
\end{APACrefauthors}%
\unskip\
\newblock
\APACrefYear{2008}.
\newblock
\APACrefbtitle {Forecasting with exponential smoothing: the state space
  approach} {Forecasting with exponential smoothing: the state space approach}.
\newblock
\APACaddressPublisher{}{Springer Science \& Business Media}.
\PrintBackRefs{\CurrentBib}

\bibitem [\protect \citeauthoryear {%
IMF%
}{%
IMF%
}{%
{\protect \APACyear {2020}}%
}]{%
imf:2020}
\APACinsertmetastar {%
imf:2020}%
\begin{APACrefauthors}%
IMF.%
\end{APACrefauthors}%
\unskip\
\newblock
\APACrefYearMonthDay{2020}{}{}.
\newblock
\APACrefbtitle {{Publication of Financial Sector Assessment Program
  Documentation - Technical Note on Financial Stability Analysis, Stress
  Testing, and Interconnectedness}.} {{Publication of Financial Sector
  Assessment Program Documentation - Technical Note on Financial Stability
  Analysis, Stress Testing, and Interconnectedness}.}
\newblock
\begin{APACrefURL}
  [{2020-11-06}]\url{https://www.imf.org/~/media/Files/Publications/CR/2020/English/1AUTEA2020006.ashx}
  \end{APACrefURL}
\PrintBackRefs{\CurrentBib}

\bibitem [\protect \citeauthoryear {%
Ishwaran%
\ \BBA {} Rao%
}{%
Ishwaran%
\ \BBA {} Rao%
}{%
{\protect \APACyear {2005}}%
}]{%
ishwaran:2005}
\APACinsertmetastar {%
ishwaran:2005}%
\begin{APACrefauthors}%
Ishwaran, H.%
\BCBT {}\ \BBA {} Rao, J\BPBI S.%
\end{APACrefauthors}%
\unskip\
\newblock
\APACrefYearMonthDay{2005}{}{}.
\newblock
{\BBOQ}\APACrefatitle {{Spike and slab variable selection: frequentist and
  Bayesian strategies}} {{Spike and slab variable selection: frequentist and
  Bayesian strategies}}.{\BBCQ}
\newblock
\APACjournalVolNumPages{Annals of statistics}{33}{2}{730--773}.
\PrintBackRefs{\CurrentBib}

\bibitem [\protect \citeauthoryear {%
Jacobs%
}{%
Jacobs%
}{%
{\protect \APACyear {2018}}%
}]{%
jacobs:2018}
\APACinsertmetastar {%
jacobs:2018}%
\begin{APACrefauthors}%
Jacobs, M.%
\end{APACrefauthors}%
\unskip\
\newblock
\APACrefYearMonthDay{2018}{}{}.
\newblock
{\BBOQ}\APACrefatitle {The validation of machine-learning models for the stress
  testing of credit risk} {The validation of machine-learning models for the
  stress testing of credit risk}.{\BBCQ}
\newblock
\APACjournalVolNumPages{Journal of Risk Management in Financial
  Institutions}{11}{3}{218--243}.
\PrintBackRefs{\CurrentBib}

\bibitem [\protect \citeauthoryear {%
Karatzoglou%
}{%
Karatzoglou%
}{%
{\protect \APACyear {2006}}%
}]{%
karatzoglou:2006}
\APACinsertmetastar {%
karatzoglou:2006}%
\begin{APACrefauthors}%
Karatzoglou, A.%
\end{APACrefauthors}%
\unskip\
\newblock
\APACrefYear{2006}.
\unskip\
\newblock
\APACrefbtitle {Kernel methods software, algorithms and applications} {Kernel
  methods software, algorithms and applications}\ \APACtypeAddressSchool
  {\BUPhD}{}{}.
\PrintBackRefs{\CurrentBib}

\bibitem [\protect \citeauthoryear {%
Keeton%
\ \BBA {} Morris%
}{%
Keeton%
\ \BBA {} Morris%
}{%
{\protect \APACyear {1987}}%
}]{%
keeton:1987}
\APACinsertmetastar {%
keeton:1987}%
\begin{APACrefauthors}%
Keeton, W\BPBI R.%
\BCBT {}\ \BBA {} Morris, C\BPBI S.%
\end{APACrefauthors}%
\unskip\
\newblock
\APACrefYearMonthDay{1987}{}{}.
\newblock
{\BBOQ}\APACrefatitle {Why do banks’ loan losses differ} {Why do banks’
  loan losses differ}.{\BBCQ}
\newblock
\APACjournalVolNumPages{Economic review}{72}{5}{3--21}.
\PrintBackRefs{\CurrentBib}

\bibitem [\protect \citeauthoryear {%
Kerbl%
\ \BBA {} Sigmund%
}{%
Kerbl%
\ \BBA {} Sigmund%
}{%
{\protect \APACyear {2011}}%
}]{%
kerbl:2011}
\APACinsertmetastar {%
kerbl:2011}%
\begin{APACrefauthors}%
Kerbl, S.%
\BCBT {}\ \BBA {} Sigmund, M.%
\end{APACrefauthors}%
\unskip\
\newblock
\APACrefYearMonthDay{2011}{}{}.
\newblock
{\BBOQ}\APACrefatitle {{What Drives Aggregate Credit Risk?}} {{What Drives
  Aggregate Credit Risk?}}{\BBCQ}
\newblock
\APACjournalVolNumPages{Oesterreichische Nationalbank Financial Stability
  Report}{22}{}{}.
\PrintBackRefs{\CurrentBib}

\bibitem [\protect \citeauthoryear {%
Koning%
, Franses%
, Hibon%
\BCBL {}\ \BBA {} Stekler%
}{%
Koning%
\ \protect \BOthers {.}}{%
{\protect \APACyear {2005}}%
}]{%
koning:2005}
\APACinsertmetastar {%
koning:2005}%
\begin{APACrefauthors}%
Koning, A\BPBI J.%
, Franses, P\BPBI H.%
, Hibon, M.%
\BCBL {}\ \BBA {} Stekler, H\BPBI O.%
\end{APACrefauthors}%
\unskip\
\newblock
\APACrefYearMonthDay{2005}{}{}.
\newblock
{\BBOQ}\APACrefatitle {{The M3 competition: Statistical tests of the results}}
  {{The M3 competition: Statistical tests of the results}}.{\BBCQ}
\newblock
\APACjournalVolNumPages{International Journal of Forecasting}{21}{3}{397--409}.
\PrintBackRefs{\CurrentBib}

\bibitem [\protect \citeauthoryear {%
Koopman%
\ \BBA {} Lucas%
}{%
Koopman%
\ \BBA {} Lucas%
}{%
{\protect \APACyear {2005}}%
}]{%
koopman:2005}
\APACinsertmetastar {%
koopman:2005}%
\begin{APACrefauthors}%
Koopman, S\BPBI J.%
\BCBT {}\ \BBA {} Lucas, A.%
\end{APACrefauthors}%
\unskip\
\newblock
\APACrefYearMonthDay{2005}{}{}.
\newblock
{\BBOQ}\APACrefatitle {Business and default cycles for credit risk} {Business
  and default cycles for credit risk}.{\BBCQ}
\newblock
\APACjournalVolNumPages{Journal of Applied Econometrics}{20}{2}{311--323}.
\PrintBackRefs{\CurrentBib}

\bibitem [\protect \citeauthoryear {%
Kuhn%
}{%
Kuhn%
}{%
{\protect \APACyear {2020}}%
}]{%
kuhn:2020}
\APACinsertmetastar {%
kuhn:2020}%
\begin{APACrefauthors}%
Kuhn, M.%
\end{APACrefauthors}%
\unskip\
\newblock
\APACrefYearMonthDay{2020}{}{}.
\newblock
{\BBOQ}\APACrefatitle {{caret: Classification and Regression Training}}
  {{caret: Classification and Regression Training}}{\BBCQ}\
  [\bibcomputersoftwaremanual].
\newblock
\begin{APACrefURL} \url{https://CRAN.R-project.org/package=caret}
  \end{APACrefURL}
\newblock
\APACrefnote{R package version 6.0-86}
\PrintBackRefs{\CurrentBib}

\bibitem [\protect \citeauthoryear {%
Kwiatkowski%
, Phillips%
, Schmidt%
\BCBL {}\ \BBA {} Shin%
}{%
Kwiatkowski%
\ \protect \BOthers {.}}{%
{\protect \APACyear {1992}}%
}]{%
kwiatkowski:1992}
\APACinsertmetastar {%
kwiatkowski:1992}%
\begin{APACrefauthors}%
Kwiatkowski, D.%
, Phillips, P.%
, Schmidt, P.%
\BCBL {}\ \BBA {} Shin, Y.%
\end{APACrefauthors}%
\unskip\
\newblock
\APACrefYearMonthDay{1992}{}{}.
\newblock
{\BBOQ}\APACrefatitle {Testing the null hypothesis of stationarity against the
  alternative of a unit root: How sure are we that economic time series have a
  unit root?} {Testing the null hypothesis of stationarity against the
  alternative of a unit root: How sure are we that economic time series have a
  unit root?}{\BBCQ}
\newblock
\APACjournalVolNumPages{Journal of econometrics}{54}{1-3}{159--178}.
\PrintBackRefs{\CurrentBib}

\bibitem [\protect \citeauthoryear {%
Leo%
, Sharma%
\BCBL {}\ \BBA {} Maddulety%
}{%
Leo%
\ \protect \BOthers {.}}{%
{\protect \APACyear {2019}}%
}]{%
leo:2019}
\APACinsertmetastar {%
leo:2019}%
\begin{APACrefauthors}%
Leo, M.%
, Sharma, S.%
\BCBL {}\ \BBA {} Maddulety, K.%
\end{APACrefauthors}%
\unskip\
\newblock
\APACrefYearMonthDay{2019}{}{}.
\newblock
{\BBOQ}\APACrefatitle {Machine learning in banking risk management: A
  literature review} {Machine learning in banking risk management: A literature
  review}.{\BBCQ}
\newblock
\APACjournalVolNumPages{Risks}{7}{1}{29}.
\PrintBackRefs{\CurrentBib}

\bibitem [\protect \citeauthoryear {%
Lukacs%
, Burnham%
\BCBL {}\ \BBA {} Anderson%
}{%
Lukacs%
\ \protect \BOthers {.}}{%
{\protect \APACyear {2010}}%
}]{%
lukacs:2010}
\APACinsertmetastar {%
lukacs:2010}%
\begin{APACrefauthors}%
Lukacs, P.%
, Burnham, K.%
\BCBL {}\ \BBA {} Anderson, D.%
\end{APACrefauthors}%
\unskip\
\newblock
\APACrefYearMonthDay{2010}{}{}.
\newblock
{\BBOQ}\APACrefatitle {{Model selection bias and Freedman’s paradox}} {{Model
  selection bias and Freedman’s paradox}}.{\BBCQ}
\newblock
\APACjournalVolNumPages{Annals of the Institute of Statistical
  Mathematics}{62}{1}{117}.
\PrintBackRefs{\CurrentBib}

\bibitem [\protect \citeauthoryear {%
Mariano%
\ \BBA {} Preve%
}{%
Mariano%
\ \BBA {} Preve%
}{%
{\protect \APACyear {2012}}%
}]{%
mariano:2012}
\APACinsertmetastar {%
mariano:2012}%
\begin{APACrefauthors}%
Mariano, R\BPBI S.%
\BCBT {}\ \BBA {} Preve, D.%
\end{APACrefauthors}%
\unskip\
\newblock
\APACrefYearMonthDay{2012}{}{}.
\newblock
{\BBOQ}\APACrefatitle {Statistical tests for multiple forecast comparison}
  {Statistical tests for multiple forecast comparison}.{\BBCQ}
\newblock
\APACjournalVolNumPages{Journal of econometrics}{169}{1}{123--130}.
\PrintBackRefs{\CurrentBib}

\bibitem [\protect \citeauthoryear {%
McCulloch%
\ \BBA {} Pitts%
}{%
McCulloch%
\ \BBA {} Pitts%
}{%
{\protect \APACyear {1943}}%
}]{%
mcculloch:1943}
\APACinsertmetastar {%
mcculloch:1943}%
\begin{APACrefauthors}%
McCulloch, W.%
\BCBT {}\ \BBA {} Pitts, W.%
\end{APACrefauthors}%
\unskip\
\newblock
\APACrefYearMonthDay{1943}{}{}.
\newblock
{\BBOQ}\APACrefatitle {A logical calculus of the ideas immanent in nervous
  activity} {A logical calculus of the ideas immanent in nervous
  activity}.{\BBCQ}
\newblock
\APACjournalVolNumPages{The bulletin of mathematical
  biophysics}{5}{4}{115--133}.
\PrintBackRefs{\CurrentBib}

\bibitem [\protect \citeauthoryear {%
Meinshausen%
}{%
Meinshausen%
}{%
{\protect \APACyear {2007}}%
}]{%
meinshausen:2007}
\APACinsertmetastar {%
meinshausen:2007}%
\begin{APACrefauthors}%
Meinshausen, N.%
\end{APACrefauthors}%
\unskip\
\newblock
\APACrefYearMonthDay{2007}{}{}.
\newblock
{\BBOQ}\APACrefatitle {Relaxed lasso} {Relaxed lasso}.{\BBCQ}
\newblock
\APACjournalVolNumPages{Computational Statistics \& Data
  Analysis}{52}{1}{374--393}.
\PrintBackRefs{\CurrentBib}

\bibitem [\protect \citeauthoryear {%
Meinshausen%
}{%
Meinshausen%
}{%
{\protect \APACyear {2010}}%
}]{%
meinshausen:2010}
\APACinsertmetastar {%
meinshausen:2010}%
\begin{APACrefauthors}%
Meinshausen, N.%
\end{APACrefauthors}%
\unskip\
\newblock
\APACrefYearMonthDay{2010}{}{}.
\newblock
{\BBOQ}\APACrefatitle {Node harvest} {Node harvest}.{\BBCQ}
\newblock
\APACjournalVolNumPages{The Annals of Applied Statistics}{}{}{2049--2072}.
\PrintBackRefs{\CurrentBib}

\bibitem [\protect \citeauthoryear {%
Meinshausen%
\ \BBA {} B{\"u}hlmann%
}{%
Meinshausen%
\ \BBA {} B{\"u}hlmann%
}{%
{\protect \APACyear {2006}}%
}]{%
meinshausen:2006}
\APACinsertmetastar {%
meinshausen:2006}%
\begin{APACrefauthors}%
Meinshausen, N.%
\BCBT {}\ \BBA {} B{\"u}hlmann, P.%
\end{APACrefauthors}%
\unskip\
\newblock
\APACrefYearMonthDay{2006}{}{}.
\newblock
{\BBOQ}\APACrefatitle {Variable selection and high-dimensional graphs with the
  lasso} {Variable selection and high-dimensional graphs with the
  lasso}.{\BBCQ}
\newblock
\APACjournalVolNumPages{Ann Stat}{34}{}{1436--1462}.
\PrintBackRefs{\CurrentBib}

\bibitem [\protect \citeauthoryear {%
Murray%
}{%
Murray%
}{%
{\protect \APACyear {2017}}%
}]{%
murray:2017}
\APACinsertmetastar {%
murray:2017}%
\begin{APACrefauthors}%
Murray, J\BPBI S.%
\end{APACrefauthors}%
\unskip\
\newblock
\APACrefYearMonthDay{2017}{}{}.
\newblock
{\BBOQ}\APACrefatitle {Log-linear Bayesian additive regression trees for
  categorical and count responses} {Log-linear bayesian additive regression
  trees for categorical and count responses}.{\BBCQ}
\newblock
\APACjournalVolNumPages{arXiv preprint arXiv:1701.01503}{3}{}{}.
\PrintBackRefs{\CurrentBib}

\bibitem [\protect \citeauthoryear {%
Nemenyi%
}{%
Nemenyi%
}{%
{\protect \APACyear {1963}}%
}]{%
nemenyi:1963}
\APACinsertmetastar {%
nemenyi:1963}%
\begin{APACrefauthors}%
Nemenyi, P\BPBI B.%
\end{APACrefauthors}%
\unskip\
\newblock
\APACrefYear{1963}.
\newblock
\APACrefbtitle {Distribution-free multiple comparisons.} {Distribution-free
  multiple comparisons.}
\newblock
\APACaddressPublisher{}{Princeton University}.
\PrintBackRefs{\CurrentBib}

\bibitem [\protect \citeauthoryear {%
Newbold%
\ \BBA {} Granger%
}{%
Newbold%
\ \BBA {} Granger%
}{%
{\protect \APACyear {1974}}%
}]{%
newbold:1974}
\APACinsertmetastar {%
newbold:1974}%
\begin{APACrefauthors}%
Newbold, P.%
\BCBT {}\ \BBA {} Granger, C\BPBI W.%
\end{APACrefauthors}%
\unskip\
\newblock
\APACrefYearMonthDay{1974}{}{}.
\newblock
{\BBOQ}\APACrefatitle {Experience with forecasting univariate time series and
  the combination of forecasts} {Experience with forecasting univariate time
  series and the combination of forecasts}.{\BBCQ}
\newblock
\APACjournalVolNumPages{Journal of the Royal Statistical Society: Series A
  (General)}{137}{2}{131--146}.
\PrintBackRefs{\CurrentBib}

\bibitem [\protect \citeauthoryear {%
Olson%
, Cava%
, Mustahsan%
, Varik%
\BCBL {}\ \BBA {} Moore%
}{%
Olson%
\ \protect \BOthers {.}}{%
{\protect \APACyear {2018}}%
}]{%
olson:2018}
\APACinsertmetastar {%
olson:2018}%
\begin{APACrefauthors}%
Olson, R\BPBI S.%
, Cava, W\BPBI L.%
, Mustahsan, Z.%
, Varik, A.%
\BCBL {}\ \BBA {} Moore, J\BPBI H.%
\end{APACrefauthors}%
\unskip\
\newblock
\APACrefYearMonthDay{2018}{}{}.
\newblock
{\BBOQ}\APACrefatitle {Data-driven advice for applying machine learning to
  bioinformatics problems} {Data-driven advice for applying machine learning to
  bioinformatics problems}.{\BBCQ}
\newblock
\BIn{} \APACrefbtitle {PACIFIC SYMPOSIUM ON BIOCOMPUTING 2018: Proceedings of
  the Pacific Symposium} {Pacific symposium on biocomputing 2018: Proceedings
  of the pacific symposium}\ (\BPGS\ 192--203).
\PrintBackRefs{\CurrentBib}

\bibitem [\protect \citeauthoryear {%
Papadopoulos%
, Papadopoulos%
\BCBL {}\ \BBA {} Sager%
}{%
Papadopoulos%
\ \protect \BOthers {.}}{%
{\protect \APACyear {2016}}%
}]{%
papadopoulos:2016}
\APACinsertmetastar {%
papadopoulos:2016}%
\begin{APACrefauthors}%
Papadopoulos, G.%
, Papadopoulos, S.%
\BCBL {}\ \BBA {} Sager, T.%
\end{APACrefauthors}%
\unskip\
\newblock
\APACrefYearMonthDay{2016}{}{}.
\newblock
\APACrefbtitle {Credit risk stress testing for EU15 banks: a model combination
  approach} {Credit risk stress testing for eu15 banks: a model combination
  approach}\ \APACbVolEdTR{}{\BTR{}}.
\PrintBackRefs{\CurrentBib}

\bibitem [\protect \citeauthoryear {%
Park%
\ \BBA {} Casella%
}{%
Park%
\ \BBA {} Casella%
}{%
{\protect \APACyear {2008}}%
}]{%
park:2008}
\APACinsertmetastar {%
park:2008}%
\begin{APACrefauthors}%
Park, T.%
\BCBT {}\ \BBA {} Casella, G.%
\end{APACrefauthors}%
\unskip\
\newblock
\APACrefYearMonthDay{2008}{}{}.
\newblock
{\BBOQ}\APACrefatitle {The bayesian lasso} {The bayesian lasso}.{\BBCQ}
\newblock
\APACjournalVolNumPages{Journal of the American Statistical
  Association}{103}{482}{681--686}.
\PrintBackRefs{\CurrentBib}

\bibitem [\protect \citeauthoryear {%
Pesaran%
, Schuermann%
, Treutler%
\BCBL {}\ \BBA {} Weiner%
}{%
Pesaran%
\ \protect \BOthers {.}}{%
{\protect \APACyear {2006}}%
}]{%
pesaran:2006}
\APACinsertmetastar {%
pesaran:2006}%
\begin{APACrefauthors}%
Pesaran, M\BPBI H.%
, Schuermann, T.%
, Treutler, B\BHBI J.%
\BCBL {}\ \BBA {} Weiner, S\BPBI M.%
\end{APACrefauthors}%
\unskip\
\newblock
\APACrefYearMonthDay{2006}{}{}.
\newblock
{\BBOQ}\APACrefatitle {Macroeconomic dynamics and credit risk: a global
  perspective} {Macroeconomic dynamics and credit risk: a global
  perspective}.{\BBCQ}
\newblock
\APACjournalVolNumPages{Journal of Money, Credit and Banking}{}{}{1211--1261}.
\PrintBackRefs{\CurrentBib}

\bibitem [\protect \citeauthoryear {%
Pesola%
}{%
Pesola%
}{%
{\protect \APACyear {2001}}%
}]{%
pesola:2001}
\APACinsertmetastar {%
pesola:2001}%
\begin{APACrefauthors}%
Pesola, J.%
\end{APACrefauthors}%
\unskip\
\newblock
\APACrefYearMonthDay{2001}{}{}.
\newblock
{\BBOQ}\APACrefatitle {The role of macroeconomic shocks in banking crises} {The
  role of macroeconomic shocks in banking crises}.{\BBCQ}
\newblock
\APACjournalVolNumPages{Bank of Finland discussion paper}{}{6}{}.
\PrintBackRefs{\CurrentBib}

\bibitem [\protect \citeauthoryear {%
D.~Phillips%
}{%
D.~Phillips%
}{%
{\protect \APACyear {1962}}%
}]{%
phillips:1962}
\APACinsertmetastar {%
phillips:1962}%
\begin{APACrefauthors}%
Phillips, D.%
\end{APACrefauthors}%
\unskip\
\newblock
\APACrefYearMonthDay{1962}{}{}.
\newblock
{\BBOQ}\APACrefatitle {A technique for the numerical solution of certain
  integral equations of the first kind} {A technique for the numerical solution
  of certain integral equations of the first kind}.{\BBCQ}
\newblock
\APACjournalVolNumPages{Journal of the ACM (JACM)}{9}{1}{84--97}.
\PrintBackRefs{\CurrentBib}

\bibitem [\protect \citeauthoryear {%
P.~Phillips%
\ \BBA {} Perron%
}{%
P.~Phillips%
\ \BBA {} Perron%
}{%
{\protect \APACyear {1988}}%
}]{%
phillips:1988}
\APACinsertmetastar {%
phillips:1988}%
\begin{APACrefauthors}%
Phillips, P.%
\BCBT {}\ \BBA {} Perron, P.%
\end{APACrefauthors}%
\unskip\
\newblock
\APACrefYearMonthDay{1988}{}{}.
\newblock
{\BBOQ}\APACrefatitle {Testing for a unit root in time series regression}
  {Testing for a unit root in time series regression}.{\BBCQ}
\newblock
\APACjournalVolNumPages{Biometrika}{75}{2}{335--346}.
\PrintBackRefs{\CurrentBib}

\bibitem [\protect \citeauthoryear {%
Pratola%
, Chipman%
, George%
\BCBL {}\ \BBA {} McCulloch%
}{%
Pratola%
\ \protect \BOthers {.}}{%
{\protect \APACyear {2020}}%
}]{%
pratola:2020}
\APACinsertmetastar {%
pratola:2020}%
\begin{APACrefauthors}%
Pratola, M\BPBI T.%
, Chipman, H\BPBI A.%
, George, E.%
\BCBL {}\ \BBA {} McCulloch, R.%
\end{APACrefauthors}%
\unskip\
\newblock
\APACrefYearMonthDay{2020}{}{}.
\newblock
{\BBOQ}\APACrefatitle {Heteroscedastic BART via multiplicative regression
  trees} {Heteroscedastic bart via multiplicative regression trees}.{\BBCQ}
\newblock
\APACjournalVolNumPages{Journal of Computational and Graphical
  Statistics}{29}{2}{405--417}.
\PrintBackRefs{\CurrentBib}

\bibitem [\protect \citeauthoryear {%
{R Core Team}%
}{%
{R Core Team}%
}{%
{\protect \APACyear {2020}}%
}]{%
R:2020}
\APACinsertmetastar {%
R:2020}%
\begin{APACrefauthors}%
{R Core Team}.%
\end{APACrefauthors}%
\unskip\
\newblock
\APACrefYearMonthDay{2020}{}{}.
\newblock
{\BBOQ}\APACrefatitle {{R: A Language and Environment for Statistical
  Computing}} {{R: A Language and Environment for Statistical
  Computing}}{\BBCQ}\ [\bibcomputersoftwaremanual].
\PrintBackRefs{\CurrentBib}

\bibitem [\protect \citeauthoryear {%
Raftery%
}{%
Raftery%
}{%
{\protect \APACyear {1995}}%
}]{%
raftery:1995}
\APACinsertmetastar {%
raftery:1995}%
\begin{APACrefauthors}%
Raftery, A\BPBI E.%
\end{APACrefauthors}%
\unskip\
\newblock
\APACrefYearMonthDay{1995}{}{}.
\newblock
{\BBOQ}\APACrefatitle {Bayesian model selection in social research} {Bayesian
  model selection in social research}.{\BBCQ}
\newblock
\APACjournalVolNumPages{Sociological methodology}{}{}{111--163}.
\PrintBackRefs{\CurrentBib}

\bibitem [\protect \citeauthoryear {%
Riedmiller%
}{%
Riedmiller%
}{%
{\protect \APACyear {1994}}%
}]{%
riedmiller:1994}
\APACinsertmetastar {%
riedmiller:1994}%
\begin{APACrefauthors}%
Riedmiller, M.%
\end{APACrefauthors}%
\unskip\
\newblock
\APACrefYearMonthDay{1994}{}{}.
\newblock
{\BBOQ}\APACrefatitle {Rprop-description and implementation details}
  {Rprop-description and implementation details}.{\BBCQ}
\newblock

\PrintBackRefs{\CurrentBib}

\bibitem [\protect \citeauthoryear {%
Robert%
, William%
\BCBL {}\ \BBA {} Irma%
}{%
Robert%
\ \protect \BOthers {.}}{%
{\protect \APACyear {1990}}%
}]{%
robert:1990}
\APACinsertmetastar {%
robert:1990}%
\begin{APACrefauthors}%
Robert, C.%
, William, C.%
\BCBL {}\ \BBA {} Irma, T.%
\end{APACrefauthors}%
\unskip\
\newblock
\APACrefYearMonthDay{1990}{}{}.
\newblock
{\BBOQ}\APACrefatitle {{STL: A seasonal-trend decomposition procedure based on
  loess}} {{STL: A seasonal-trend decomposition procedure based on
  loess}}.{\BBCQ}
\newblock
\APACjournalVolNumPages{Journal of official statistics}{6}{1}{3--73}.
\PrintBackRefs{\CurrentBib}

\bibitem [\protect \citeauthoryear {%
Schechtman%
\ \BBA {} Gaglianone%
}{%
Schechtman%
\ \BBA {} Gaglianone%
}{%
{\protect \APACyear {2012}}%
}]{%
schechtman:2012}
\APACinsertmetastar {%
schechtman:2012}%
\begin{APACrefauthors}%
Schechtman, R.%
\BCBT {}\ \BBA {} Gaglianone, W\BPBI P.%
\end{APACrefauthors}%
\unskip\
\newblock
\APACrefYearMonthDay{2012}{}{}.
\newblock
{\BBOQ}\APACrefatitle {Macro stress testing of credit risk focused on the
  tails} {Macro stress testing of credit risk focused on the tails}.{\BBCQ}
\newblock
\APACjournalVolNumPages{Journal of Financial Stability}{8}{3}{174--192}.
\PrintBackRefs{\CurrentBib}

\bibitem [\protect \citeauthoryear {%
Schmid%
\ \BBA {} Hothorn%
}{%
Schmid%
\ \BBA {} Hothorn%
}{%
{\protect \APACyear {2008}}%
}]{%
schmid:2008}
\APACinsertmetastar {%
schmid:2008}%
\begin{APACrefauthors}%
Schmid, M.%
\BCBT {}\ \BBA {} Hothorn, T.%
\end{APACrefauthors}%
\unskip\
\newblock
\APACrefYearMonthDay{2008}{}{}.
\newblock
{\BBOQ}\APACrefatitle {{Boosting additive models using component-wise
  P-splines}} {{Boosting additive models using component-wise
  P-splines}}.{\BBCQ}
\newblock
\APACjournalVolNumPages{Computational Statistics \& Data
  Analysis}{53}{2}{298--311}.
\PrintBackRefs{\CurrentBib}

\bibitem [\protect \citeauthoryear {%
Sparapani%
, Logan%
, McCulloch%
\BCBL {}\ \BBA {} Laud%
}{%
Sparapani%
\ \protect \BOthers {.}}{%
{\protect \APACyear {2016}}%
}]{%
sparapani:2016}
\APACinsertmetastar {%
sparapani:2016}%
\begin{APACrefauthors}%
Sparapani, R\BPBI A.%
, Logan, B\BPBI R.%
, McCulloch, R.%
\BCBL {}\ \BBA {} Laud, P\BPBI W.%
\end{APACrefauthors}%
\unskip\
\newblock
\APACrefYearMonthDay{2016}{}{}.
\newblock
{\BBOQ}\APACrefatitle {{Nonparametric survival analysis using Bayesian additive
  regression trees (BART)}} {{Nonparametric survival analysis using Bayesian
  additive regression trees (BART)}}.{\BBCQ}
\newblock
\APACjournalVolNumPages{Statistics in medicine}{35}{16}{2741--2753}.
\PrintBackRefs{\CurrentBib}

\bibitem [\protect \citeauthoryear {%
Svetunkov%
}{%
Svetunkov%
}{%
{\protect \APACyear {2016}}%
}]{%
svetunkov:2016}
\APACinsertmetastar {%
svetunkov:2016}%
\begin{APACrefauthors}%
Svetunkov, I.%
\end{APACrefauthors}%
\unskip\
\newblock
\APACrefYear{2016}.
\newblock
\APACrefbtitle {Complex exponential smoothing} {Complex exponential smoothing}.
\newblock
\APACaddressPublisher{}{Lancaster University (United Kingdom)}.
\PrintBackRefs{\CurrentBib}

\bibitem [\protect \citeauthoryear {%
Svetunkov%
}{%
Svetunkov%
}{%
{\protect \APACyear {2020}}%
}]{%
svetunkov:2020}
\APACinsertmetastar {%
svetunkov:2020}%
\begin{APACrefauthors}%
Svetunkov, I.%
\end{APACrefauthors}%
\unskip\
\newblock
\APACrefYearMonthDay{2020}{}{}.
\newblock
{\BBOQ}\APACrefatitle {{greybox: Toolbox for Model Building and Forecasting}}
  {{greybox: Toolbox for Model Building and Forecasting}}{\BBCQ}\
  [\bibcomputersoftwaremanual].
\newblock
\begin{APACrefURL} \url{https://CRAN.R-project.org/package=greybox}
  \end{APACrefURL}
\newblock
\APACrefnote{R package version 0.6.0}
\PrintBackRefs{\CurrentBib}

\bibitem [\protect \citeauthoryear {%
Tibshirani%
}{%
Tibshirani%
}{%
{\protect \APACyear {1996}}%
}]{%
tibshirani:1996}
\APACinsertmetastar {%
tibshirani:1996}%
\begin{APACrefauthors}%
Tibshirani, R.%
\end{APACrefauthors}%
\unskip\
\newblock
\APACrefYearMonthDay{1996}{}{}.
\newblock
{\BBOQ}\APACrefatitle {Regression shrinkage and selection via the lasso}
  {Regression shrinkage and selection via the lasso}.{\BBCQ}
\newblock
\APACjournalVolNumPages{Journal of the Royal Statistical Society: Series B
  (Methodological)}{58}{1}{267--288}.
\PrintBackRefs{\CurrentBib}

\bibitem [\protect \citeauthoryear {%
Tibshirani%
, Saunders%
, Rosset%
, Zhu%
\BCBL {}\ \BBA {} Knight%
}{%
Tibshirani%
\ \protect \BOthers {.}}{%
{\protect \APACyear {2005}}%
}]{%
tibshirani:2005}
\APACinsertmetastar {%
tibshirani:2005}%
\begin{APACrefauthors}%
Tibshirani, R.%
, Saunders, M.%
, Rosset, S.%
, Zhu, J.%
\BCBL {}\ \BBA {} Knight, K.%
\end{APACrefauthors}%
\unskip\
\newblock
\APACrefYearMonthDay{2005}{}{}.
\newblock
{\BBOQ}\APACrefatitle {Sparsity and smoothness via the fused lasso} {Sparsity
  and smoothness via the fused lasso}.{\BBCQ}
\newblock
\APACjournalVolNumPages{Journal of the Royal Statistical Society: Series B
  (Statistical Methodology)}{67}{1}{91--108}.
\PrintBackRefs{\CurrentBib}

\bibitem [\protect \citeauthoryear {%
Tikhonov%
}{%
Tikhonov%
}{%
{\protect \APACyear {1943}}%
}]{%
tikhonov:1943}
\APACinsertmetastar {%
tikhonov:1943}%
\begin{APACrefauthors}%
Tikhonov, A\BPBI N.%
\end{APACrefauthors}%
\unskip\
\newblock
\APACrefYearMonthDay{1943}{}{}.
\newblock
{\BBOQ}\APACrefatitle {On the stability of inverse problems} {On the stability
  of inverse problems}.{\BBCQ}
\newblock
\BIn{} \APACrefbtitle {Dokl. Akad. Nauk SSSR} {Dokl. akad. nauk sssr}\
  (\BVOL~39, \BPGS\ 195--198).
\PrintBackRefs{\CurrentBib}

\bibitem [\protect \citeauthoryear {%
Tipping%
}{%
Tipping%
}{%
{\protect \APACyear {2001}}%
}]{%
tipping:2001}
\APACinsertmetastar {%
tipping:2001}%
\begin{APACrefauthors}%
Tipping, M.%
\end{APACrefauthors}%
\unskip\
\newblock
\APACrefYearMonthDay{2001}{}{}.
\newblock
{\BBOQ}\APACrefatitle {Sparse Bayesian learning and the relevance vector
  machine} {Sparse bayesian learning and the relevance vector machine}.{\BBCQ}
\newblock
\APACjournalVolNumPages{Journal of machine learning
  research}{1}{Jun}{211--244}.
\PrintBackRefs{\CurrentBib}

\bibitem [\protect \citeauthoryear {%
Van~Rossum%
\ \BBA {} Drake%
}{%
Van~Rossum%
\ \BBA {} Drake%
}{%
{\protect \APACyear {2009}}%
}]{%
vanrossum:2009}
\APACinsertmetastar {%
vanrossum:2009}%
\begin{APACrefauthors}%
Van~Rossum, G.%
\BCBT {}\ \BBA {} Drake, F\BPBI L.%
\end{APACrefauthors}%
\unskip\
\newblock
\APACrefYear{2009}.
\newblock
\APACrefbtitle {{Python 3 Reference Manual}} {{Python 3 Reference Manual}}.
\newblock
\APACaddressPublisher{Scotts Valley, CA}{CreateSpace}.
\PrintBackRefs{\CurrentBib}

\bibitem [\protect \citeauthoryear {%
Weiss%
, Raviv%
\BCBL {}\ \BBA {} Roetzer%
}{%
Weiss%
\ \protect \BOthers {.}}{%
{\protect \APACyear {2018}}%
}]{%
weiss:2018b}
\APACinsertmetastar {%
weiss:2018b}%
\begin{APACrefauthors}%
Weiss, C\BPBI E.%
, Raviv, E.%
\BCBL {}\ \BBA {} Roetzer, G.%
\end{APACrefauthors}%
\unskip\
\newblock
\APACrefYearMonthDay{2018}{}{}.
\newblock
{\BBOQ}\APACrefatitle {{Forecast Combinations in R using the ForecastComb
  Package.}} {{Forecast Combinations in R using the ForecastComb
  Package.}}{\BBCQ}
\newblock
\APACjournalVolNumPages{R Journal}{10}{2}{}.
\PrintBackRefs{\CurrentBib}

\bibitem [\protect \citeauthoryear {%
Wilcoxon%
}{%
Wilcoxon%
}{%
{\protect \APACyear {1945}}%
}]{%
wilcoxon:1945}
\APACinsertmetastar {%
wilcoxon:1945}%
\begin{APACrefauthors}%
Wilcoxon, F.%
\end{APACrefauthors}%
\unskip\
\newblock
\APACrefYearMonthDay{1945}{}{}.
\newblock
{\BBOQ}\APACrefatitle {Individual Comparisons by Ranking Methods} {Individual
  comparisons by ranking methods}.{\BBCQ}
\newblock
\APACjournalVolNumPages{Biometrics Bulletin}{1}{6}{80--83}.
\PrintBackRefs{\CurrentBib}

\bibitem [\protect \citeauthoryear {%
Williams%
\ \BBA {} Rasmussen%
}{%
Williams%
\ \BBA {} Rasmussen%
}{%
{\protect \APACyear {1996}}%
}]{%
williams:1996}
\APACinsertmetastar {%
williams:1996}%
\begin{APACrefauthors}%
Williams, C.%
\BCBT {}\ \BBA {} Rasmussen, C\BPBI E.%
\end{APACrefauthors}%
\unskip\
\newblock
\APACrefYearMonthDay{1996}{}{}.
\newblock
{\BBOQ}\APACrefatitle {Gaussian processes for regression} {Gaussian processes
  for regression}.{\BBCQ}
\newblock

\PrintBackRefs{\CurrentBib}

\bibitem [\protect \citeauthoryear {%
Wilson%
}{%
Wilson%
}{%
{\protect \APACyear {1998}}%
}]{%
wilson:1998}
\APACinsertmetastar {%
wilson:1998}%
\begin{APACrefauthors}%
Wilson, T\BPBI C.%
\end{APACrefauthors}%
\unskip\
\newblock
\APACrefYearMonthDay{1998}{}{}.
\newblock
{\BBOQ}\APACrefatitle {Portfolio credit risk} {Portfolio credit risk}.{\BBCQ}
\newblock
\APACjournalVolNumPages{Economic Policy Review}{4}{3}{}.
\PrintBackRefs{\CurrentBib}

\bibitem [\protect \citeauthoryear {%
Wold%
, Sj{\"o}str{\"o}m%
\BCBL {}\ \BBA {} Eriksson%
}{%
Wold%
\ \protect \BOthers {.}}{%
{\protect \APACyear {2001}}%
}]{%
wold:2001}
\APACinsertmetastar {%
wold:2001}%
\begin{APACrefauthors}%
Wold, S.%
, Sj{\"o}str{\"o}m, M.%
\BCBL {}\ \BBA {} Eriksson, L.%
\end{APACrefauthors}%
\unskip\
\newblock
\APACrefYearMonthDay{2001}{}{}.
\newblock
{\BBOQ}\APACrefatitle {PLS-regression: a basic tool of chemometrics}
  {Pls-regression: a basic tool of chemometrics}.{\BBCQ}
\newblock
\APACjournalVolNumPages{Chemometrics and intelligent laboratory
  systems}{58}{2}{109--130}.
\PrintBackRefs{\CurrentBib}

\bibitem [\protect \citeauthoryear {%
Zou%
}{%
Zou%
}{%
{\protect \APACyear {2006}}%
}]{%
zou:2006}
\APACinsertmetastar {%
zou:2006}%
\begin{APACrefauthors}%
Zou, H.%
\end{APACrefauthors}%
\unskip\
\newblock
\APACrefYearMonthDay{2006}{}{}.
\newblock
{\BBOQ}\APACrefatitle {The adaptive lasso and its oracle properties} {The
  adaptive lasso and its oracle properties}.{\BBCQ}
\newblock
\APACjournalVolNumPages{Journal of the American statistical
  association}{101}{476}{1418--1429}.
\PrintBackRefs{\CurrentBib}

\bibitem [\protect \citeauthoryear {%
Zou%
\ \BBA {} Hastie%
}{%
Zou%
\ \BBA {} Hastie%
}{%
{\protect \APACyear {2005}}%
}]{%
zou:2005}
\APACinsertmetastar {%
zou:2005}%
\begin{APACrefauthors}%
Zou, H.%
\BCBT {}\ \BBA {} Hastie, T.%
\end{APACrefauthors}%
\unskip\
\newblock
\APACrefYearMonthDay{2005}{}{}.
\newblock
{\BBOQ}\APACrefatitle {Regularization and variable selection via the elastic
  net} {Regularization and variable selection via the elastic net}.{\BBCQ}
\newblock
\APACjournalVolNumPages{Journal of the royal statistical society: series B
  (statistical methodology)}{67}{2}{301--320}.
\PrintBackRefs{\CurrentBib}

\end{thebibliography}

\end{document}